\documentclass[aps,prc,amsfonts,preprintnumbers,superscriptaddress,showpacs,nofootinbib]{revtex4-2}
\usepackage{amssymb,amsmath}
\usepackage{mathtools}
\usepackage{color}
\usepackage[pdfencoding=auto, psdextra]{hyperref}
\usepackage{bbold}

\allowdisplaybreaks

\newcommand{\fet}[1]{\mbox{\boldmath $#1$}}

\newcommand{\beq}{\begin{equation}}
\newcommand{\eeq}{\end{equation}}
\newcommand{\beqa}{\begin{eqnarray}}
\newcommand{\eeqa}{\end{eqnarray}}

\newcommand{\ben}{\begin{displaymath}}
\newcommand{\een}{\end{displaymath}}
\newcommand{\be}{\begin{equation}}
\newcommand{\ee}{\end{equation}}
\newcommand{\bea}{\begin{eqnarray}}
\newcommand{\eea}{\end{eqnarray}}
\newcommand{\nn}{\nonumber \\ }

\begin{document}

\title{Parity and time-reversal violating nuclear forces with explicit $\Delta$-excitations}

\author{L.~Gandor}
\email[]{Email: lukas.gandor@rub.de}
\affiliation{Institut f\"ur Theoretische Physik II, Ruhr-Universit\"at Bochum,
 D-44780 Bochum, Germany}
\author{H.~Krebs}
\email[]{Email: hermann.krebs@rub.de}
\affiliation{Institut f\"ur Theoretische Physik II, Ruhr-Universit\"at Bochum,
  D-44780 Bochum, Germany}
\author{E.~Epelbaum}
\email[]{Email: evgeny.epelbaum@rub.de}
\affiliation{Institut f\"ur Theoretische Physik II, Ruhr-Universit\"at Bochum,
  D-44780 Bochum, Germany}

\begin{abstract}
We emphasize the usefulness of treating delta resonances as explicit
degrees of freedom in applications of chiral effective field theory (EFT)
to
parity-violating and time-reversal-violating (PVTV) nuclear
interactions. Compared with the delta-less framework, the explicit
inclusion of the delta isobar allows one to resum certain types of
contributions to the PVTV two-pion exchange two- and three-nucleon
potentials without at the same time introducing any unknown parameters
up to
next-to-next-to-leading order in the EFT expansion. We provide the
corresponding expressions for the delta contributions in momentum
and coordinate spaces and compare the convergence of the EFT expansion
in both formulations.
\end{abstract}


\maketitle

\vspace{-0.2cm}

\section{Introduction}
\label{sec:intro}

Parity-violating time-reversal-violating nuclear interactions play an
important role in research focused on understanding the observed
matter-antimatter asymmetry in the Universe and in searches for
physics beyond the Standard Model. This includes, in particular, 
ongoing efforts towards an experimental observation of the permanent
electric dipole moment of the nucleon and nuclei, which can only
emerge from PVTV
interactions. The CP violation in the Standard
Model, which originates from complex phases in the quark and neutrino
mixing matrices as well as from the $\theta$-term in the strong sector,
is known to be insufficient to describe the observed
matter-antimatter asymmetry. On the other hand, when viewing the
Standard Model as a low-energy approximation in the EFT framework,
further sources of CP violation appear from 
dimension-six and higher operators. In fact, many
beyond-Standard-Model extensions including, e.g., supersymmetric
models give rise to PVTV mechanisms, which are parametrized
by higher-dimensional operators in the Standard Model EFT. 

Regardless of the microscopic origin of CP violation in or beyond the
Standard Model, a theoretical analysis of PVTV observables in nuclear
systems can be most efficiently carried out in the framework of chiral
EFT, which is tailored to describe phenomena at momentum scales
relevant for nuclear physics. Here, the starting point is the most
general effective Lagrangian written in terms of the relevant hadronic
degrees of freedom, which respects the symmetries of the Standard
Model (or Standard Model EFT), such as especially the spontaneously
broken approximate chiral symmetry of QCD. CP violation can be
incorporated in the effective chiral Lagrangian by introducing the
corresponding terms, whose coupling constants are ultimately driven by
microscopic mechanisms of CP violation and thus expected to be very
small. The dominant CP violating interactions between pions and
nucleons are parametrized in terms of four dimensionless low-energy
constants (LECs) \cite{Mereghetti:2010tp,deVries:2012ab,Bsaisou:2014zwa,Bsaisou:2014oka}, which will be specified in the next section. The
corresponding derivative-less vertices in the effective chiral
Lagrangian are sufficient to determine the long-range PVTV nuclear
interactions up to next-to-next-to-leading order (N$^2$LO) in the 
chiral EFT expansion. Microscopic mechanisms of CP violation
lead to different scaling of these four LECs. Thus, by tuning these
LECs to hypothetical experimental PVTV signals in nuclear systems
one may be able to draw conclusions about specific sources of CP
violation in or beyond the Standard Model. 

PVTV nuclear forces have already been derived from the effective
chiral Lagrangian up through N$^2$LO\footnote{Throughout our paper, we
use the same terminology as in Ref.~\cite{deVries:2020iea} and treat all
dimensionless PVTV LECs as quantities on equal footing of order $\mathcal{O} (1)$,
i.e., we do not take into account suppression factors that may arise in
specific microscopic CP violating scenarios unless explicitly stated otherwise.}
\cite{Maekawa:2011vs,Bsaisou:2012rg,Bsaisou:2014zwa} and applied to
study 
selected PVTV nuclear processes \cite{Bsaisou:2014zwa,Gnech:2019dod}. 
PVTV nuclear current operators have been discussed in
Refs.~\cite{Bsaisou:2012rg,deVries:2011an}. For a review of PVTV
nuclear forces and currents and applications to nuclear systems see
Ref.~\cite{deVries:2020iea}. However, all these studies are based on
the formulation of chiral EFT in terms of pions and nucleons as the
only degrees of freedom in the effective chiral Lagrangian. On
the other hand, it is well known from studies of the usual parity-conserving
time-reversal-conserving (PCTC) nuclear interactions that such a
framework may suffer from convergence issues at low orders in the EFT
expansion due to the implicit treatment of the $\Delta$(1232)
resonance
\cite{Ordonez:1995rz,Kaiser:1998wa,Krebs:2007rh,Epelbaum:2007sq,Epelbaum:2008td,Krebs:2018jkc},
which leads to large numerical values of some of the
subleading pion-nucleon LECs. In this paper, we work out the long-range
contributions to the nuclear forces mediated by intermediate
delta-excitations  up to N$^2$LO using the formulation of chiral EFT with
explicit delta-isobar degrees of freedom. As discussed in the next
section, the explicit inclusion of the delta-isobars does not lead to
additional PVTV LECs, thereby allowing for a resummation of dominant
higher-order contributions in a parameter-free fashion.  

Our paper is organized as follows. In Sec.~\ref{sec:MomSpace}, we
specify the effective Lagrangian relevant for our calculations,
discuss the power counting and provide expressions for the
renormalized two- (2N) and three-nucleon (3N) potentials up to N$^2$LO. The
coordinate-space expressions for various long-range contributions are
given in Sec.~\ref{sec:CoordSpace}, where we also compare the
convergence of the delta-less and delta-full formulations of
chiral EFT. The main results of our study are briefly summarized in
Sec.~\ref{sec:Summary}.

\section{PVTV potentials in momentum space}
\label{sec:MomSpace}

Both parity conserving and PVTV contributions to the effective chiral
Lagrangian are well documented in the literature, see Refs.~\cite{Gasser:1983yg,Bernard:1995dp,Fettes:2000gb,Hemmert:1997ye}
and \cite{Mereghetti:2010tp,Bsaisou:2014oka}, respectively. We
therefore only list below terms relevant for our calculation, which emerge from
expanding the covariantly transforming building blocks of the
effective Lagrangian in powers of the pion fields:
\beqa
\label{Lagr}
\mathcal{L}_{\pi\pi}^{\Delta_i=0} &=& \frac{1}{2} \partial_\mu \fet \pi
\cdot \partial^\mu \fet \pi - \frac{1}{2} M^2 \fet \pi^2 +
\ldots\,,\nn
\mathcal{L}_{\pi N}^{\Delta_i=0} &=& N^\dagger \bigg[ i v \cdot
\partial - \frac{1}{4 F^2} \fet \tau \times \fet \pi \cdot (v \cdot
\partial \fet \pi ) - \frac{\mathring{g}_A}{F} \fet \tau \cdot (S \cdot
\partial \fet \pi) \bigg] N + \ldots\,, \nn
\mathcal{L}_{\pi N}^{\Delta_i=1} &=&
N^\dagger \bigg[- \frac{2 c_1}{F^2} M^2 \fet \pi^2  +
\frac{c_2}{F^2}  ( v \cdot \partial \fet \pi) \cdot (v \cdot
\partial\fet \pi ) + \frac{c_3}{F^2}
(\partial _\mu \fet \pi ) \cdot (\partial^\mu \fet \pi )
- \frac{ic_4}{F^2} \Big[ S_\mu , \; S_\nu \Big]  \fet \tau \times
(\partial^\nu \fet \pi) \cdot (\partial^\mu \fet \pi )
\bigg] N \nn
&+& \ldots\,, \nn
\mathcal{L}_{\pi N\Delta}^{\Delta_i=0} &=& -
\frac{\mathring{h}_A}{F}N^\dagger \fet T_\mu \cdot \partial^\mu \fet
\pi + {\rm H.c.} + \ldots\,, \nn
\mathcal{L}_{\pi\pi, \; {\rm PVTV}}^{\Delta_i=-2} &=& \Delta_3 m_N
\pi_3 \fet \pi^2 + \ldots\,,\nn
\mathcal{L}_{\pi N, \; {\rm PVTV}}^{\Delta_i=-1} &=&
N^\dagger \big[ g_0 \fet \tau \cdot \fet \pi + g_1 \pi_3 + g_2 \pi_3
\tau_3 \big] N + \ldots\,,
\eeqa
where the ellipses refer to terms involving a larger number of pion
fields $\fet \pi$, which are not relevant for this work.  
Here, $N$ and $\fet T_\mu$
denote the large components of the nucleon and the $\Delta$ field in
the Rarita-Schwinger formalism,
respectively, which depend on the four-velocity $v$. Next, $\fet
\tau$ denote the isospin Pauli matrices, $S_\mu = -1/4 \gamma_5
[\gamma_\mu, \gamma_\nu ] v^\nu$ is the covariant spin operator of the
nucleon, $M$ is the pion mass to leading order in quark masses while
$F$, $\mathring{g}_A$ and $\mathring{h}_A$ are the chiral-limit
values of the pion decay
constant, the nucleon and $\pi N \Delta$ axial couplings,
respectively. Further, $c_i$, $i=1, \ldots , 4$ are LECs accompanying
the  subleading pion-nucleon vertices,
while $\Delta_3$ and $g_i$ with $i=0, 1, 2$ are the LECs of the
lowest-order pionic and pion-nucleon PVTV vertices. The appearance of
the nucleon mass $m_N$ in $\mathcal{L}_{\pi\pi, \; {\rm
    PVTV}}^{\Delta_i=-2}$ is just a matter of convention for keeping
the PVTV LEC $\Delta_3$ dimensionless.

The superscripts of the
effective Lagrangians in Eq.~(\ref{Lagr}) refer to the vertex dimension
$\Delta_i$ introduced in Ref.~\cite{Weinberg:1990rz} and defined as
\beq
\Delta_i = \frac{n_i}{2} + d_i - 2\,, 
\eeq
where $n_i$ is the number of baryon field operators and $d_i$ is the
number of derivatives
or pion mass insertion at a vertex of type $i$. The EFT order $Q^\nu$ of a
connected  $N$-nucleon irreducible\footnote{Here and in what follows, the term
  irreducible refers to time-ordered diagrams that do not involve
  purely nucleonic intermediate states. Such diagrams are free of 
  infrared divergences in the $m_N \to \infty$ limit, and their
  contributions obey the standard power counting of
  chiral perturbation theory \cite{Weinberg:1990rz}.} diagram with $L$
loops made out of $V_i$ vertices of type $i$ is then given by \cite{Weinberg:1990rz,Epelbaum:2007us}
\beq
\nu = -4 + 2 (N+L) + \sum_i V_i \Delta_i \,.
\eeq
The expansion parameter of chiral EFT is given by $Q \in \{| \vec p \, |/\Lambda_b , \; M_\pi/\Lambda_b
\}$, with $M_\pi$
referring to the pion mass, $| \vec p \, |\sim M_\pi$ to generic momenta of
the nucleon and $\Lambda_b$ to the breakdown scale of the chiral
EFT expansion. Throughout this work, we employ our usual counting for the nucleon
mass with $m_N \sim \mathcal{O} (\Lambda_b^2/M_\pi ) \gg \Lambda_b$
\cite{Weinberg:1990rz,Epelbaum:2008ga}, which is different from the
assignment $m_N  \sim \mathcal{O} (\Lambda_b )$  adopted in
Ref.~\cite{Gnech:2019dod}. Notice that we treat the LECs associated 
with leading-order (LO) three-pion vertex as $\Delta_3 m_N \sim
\mathcal{O} (1)$\footnote{In our counting scheme, it would be more
  appropriate to express this LEC in terms of the dimensionless
  coupling $\Delta_3$ as $\Delta_3 \Lambda_b$, but we
  decided to keep the established notation.}.

While parity-violating time-reversal-conserving interactions has
already been considered in chiral EFT with explicit delta degrees of
freedom \cite{Kaiser:2007zzb}, PVTV effective chiral Lagrangian involving
$\Delta$ isobars has, to the best of our knowledge, not been worked out yet.

However, it is easy to see
that the lowest possible PVTV $\pi N \Delta$ terms must include at
least one derivative in order to maintain Lorentz invariance. In the
Rarita-Schwinger formulation, the $\Delta$ field appears in
combination with the spin-$3/2$ projection operator
\beq
P_{\mu \nu} = g_{\mu \nu} - v_\mu v_\nu - \frac{4}{1-d} S_\mu S_\nu\,,
\eeq
where $g_{\mu \nu}$ is the Minkowski metric tensor and 
$d$ the number of space-time dimensions. The
relationships $v^2 = 1$, $v \cdot S = 0$ and $S^2 = (1-d)/4$ imply
that both possible structures $N^\dagger v^\mu P_{\mu \nu} \fet T^\nu$
and $N^\dagger S^\mu P_{\mu \nu} \fet T^\nu$ vanish, and the only
way to obtain a nonvanishing Lorentz-invariant $\pi N \Delta$ term is by contracting
$P_{\mu \nu} \fet T^\nu$ with a four-derivative. The argument becomes even
more transparent if the Lagrangian is written directly in terms of a
four-component spin-$3/2$ field as done e.g.~in
Ref.~\cite{Kaiser:1998wa}. This requires the introduction of the $2
\times 4$ spin transition matrix $\vec S$, which has to be contracted
with a spatial derivative in order to restore rotational
invariance. Thus, the lowest-order PVTV  $\pi N \Delta$ vertices have
$\Delta_i = 0$ and start contributing to the nuclear forces at
next-to-next-to-next-to-leading  order $Q^2$ (N$^3$LO), which is beyond the
accuracy level of our calculation.

\begin{figure}[tb]
\includegraphics[width=\textwidth]{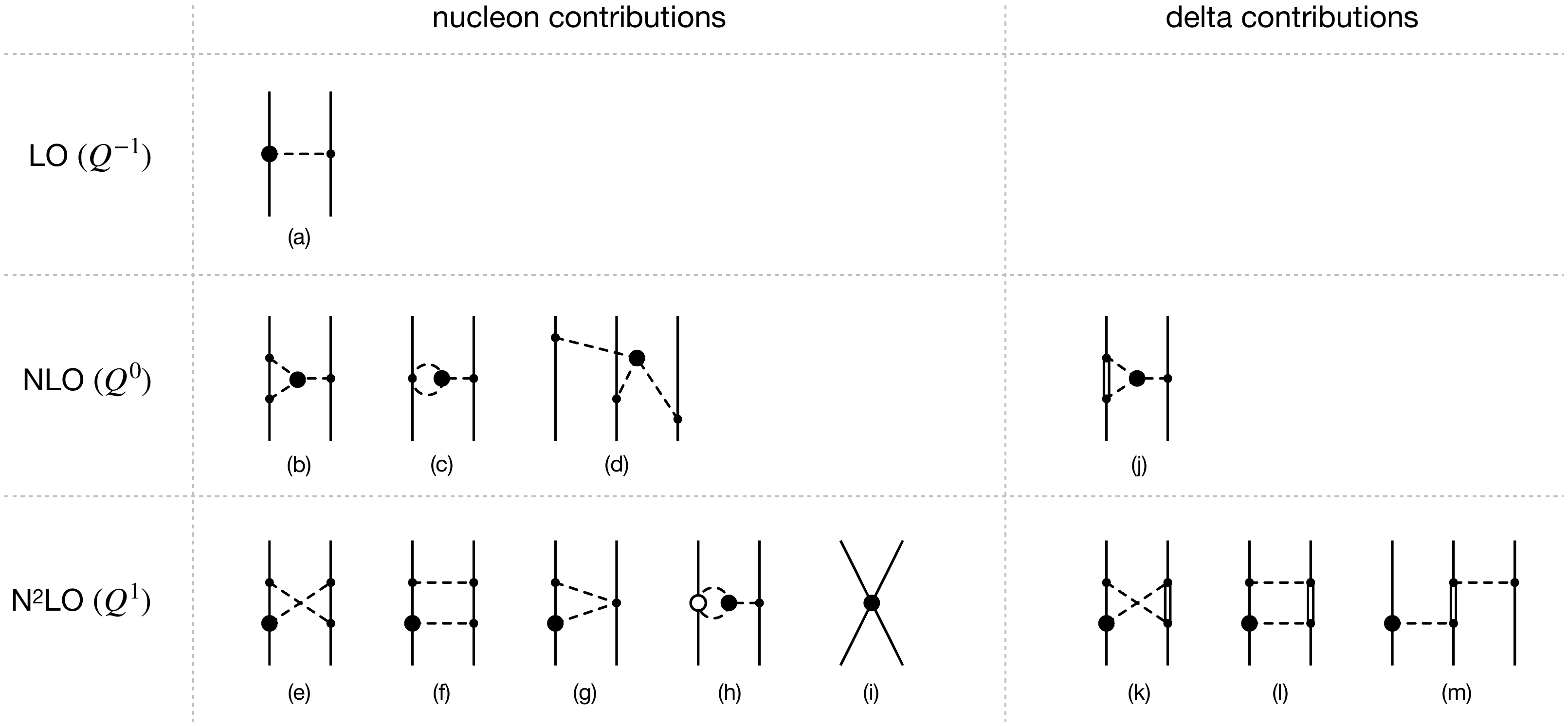}\\
  \caption{Diagrams contributing to the PVTV
    nuclear potentials at first three orders in the chiral EFT
    expansion. Solid, dashed and double lines refer to nucleons, pions
    and deltas, respectively. Solid dots (open circles) denote the
    lowest-order (subleading) PCTC
    vertices with $\Delta_i = 0$ ($\Delta_i = 1$). Filled circles
    refer to the leading PVTV pion, pion-nucleon and two-nucleon
    vertices with $\Delta_i = -2$, $\Delta_i = -1$ and $\Delta_i = 1$,
    respectively. Diagrams resulting from interchanging the order of
    vertices at the nucleon lines and those leading solely to
    renormalization of various LECs are not shown. We also  do
    not show three-nucleon diagrams made out of the LO vertices that
    can be drawn at N$^2$LO but lead to vanishing contributions
    (in a complete analogy with the vanishing NLO PCTC three-nucleon forces).}
\label{fig1}
\end{figure}

In Fig.~\ref{fig1} we show various diagrams that contribute to the
nuclear potentials up to N$^2$LO. Nuclear forces are defined in terms
of irreducible contributions to the scattering amplitude, which cannot
be generated through iterations of the Lippmann-Schwinger equation
\cite{Weinberg:1990rz}. Their derivation from the effective chiral Lagrangian can be
achieved using a variety of methods including time-ordered
perturbation theory
\cite{Weinberg:1990rz,Ordonez:1995rz,Gnech:2019dod},
S-matrix matching \cite{Kaiser:1998wa,Kaiser:1997mw}, the method of unitary
transformation \cite{Epelbaum:1998ka,Epelbaum:2007us} and the path
integral approach \cite{Krebs:2023ljo}, see
Refs.~\cite{Epelbaum:2005pn,Epelbaum:2019kcf} for a discussion of
various approaches. On the other hand, none of the diagrams shown in 
Fig.~\ref{fig1} except the box graph (f) involve reducible
time-ordered topologies, and their contributions to the nuclear
forces can therefore be obtained by calculating the corresponding Feynman
diagrams. At the accuracy level of our calculation, all methods mentioned above
lead to identical results, provided one sticks to 
energy-independent nuclear potentials. The purely nucleonic diagrams
(a)-(i) have been calculated in Ref.~\cite{Gnech:2019dod}. Notice that
differently to the counting scheme of that study, no relativistic corrections contribute to the PVTV nuclear
interactions up to N$^2$LO in our power counting scheme. We 
further remind the reader that the leading relativistic
corrections to the one-pion exchange potential  in the
energy-independent formulation are suppressed by the factor of 
$M_\pi^2/m_N^{2} \sim Q^4$ relative to the static result. Last but not least, there are numerous
diagrams at N$^2$LO that result from one-loop dressing of the PVTV and
PCTC one-pion-exchange
and the leading-order (LO) PCTC 2N contact interaction, which are not
shown in Fig.~\ref{fig1}.  The latter
type of diagrams involve the PVTV $\pi N$-vertices and lead to vanishing
N$^2$LO contributions due to the integrands being odd functions of
momenta. This also applies to graphs resulting from dressing the
PCTC one-pion exchange with PVTV nucleon-self-energy-type of
diagrams. The remaining one-loop corrections to the one-pion exchange,
except for diagrams involving the $\Delta_3$-vertex to be discussed below,
only lead to renormalization of various LECs. At the order we are
working, their contributions are accounted for by replacing $M$, $F$,
$\mathring{g}_A$ and $g_i$ with the corresponding renormalized
values $M_\pi$, $F_\pi$,
${g}_A$ and $\bar g_i$.  

Up to the order we are working, the isospin-spin-momentum structure of
the two-nucleon PVTV one- and two-pion exchange potentials can be
expressed as
\beq
\label{NotationMom}
V (\vec q \, ) = \big[ V_-^{\rm I} + W_-^{\rm I} \fet \tau_1 \cdot \fet \tau_2  +
V_-^{\rm II}  \tau_1^3
\tau_2^3 + V_-^{\rm III} (\tau_1^3 + \tau_2^3) \big] \, i (\vec q \cdot
\vec \sigma_1 - \vec q \cdot \vec \sigma_2 ) + V_+^{\rm IV} (\tau_1^3
- \tau_2^3) \, i 
 (\vec q \cdot
\vec \sigma_1 + \vec q \cdot \vec \sigma_2 ) \,,
\eeq
where $\vec \sigma_i$ ($\fet \tau_i$) denote the spin (isospin)
matrices of the nucleon $i$.  
Further, $V_-^{\rm I} (q)$,  $W_-^{\rm I} (q)$, $V_-^{\rm II} (q)$, $V_-^{\rm III} (q)$
and $V_+^{\rm IV} (q)$ are real scalar functions while $\vec q = \vec p \, ' -
\vec p$ is the momentum transfer with $\vec p$ and $\vec p \, '$
referring to the initial and final momenta in the center-of-mass frame
and $q \equiv | \vec q \, |$. The superscripts of the scalar functions
refer to the usual classification scheme of the isospin dependence of
the 2N force. Specifically, class-I, II, III and IV potentials
correspond to isospin-invariant, charge-independence breaking,
charge-symmetry breaking and isospin mixing interactions, respectively
\cite{HenleyMiller,Epelbaum:2004xf}.  

The leading-order PVTV 2N potential is generated by the one-pion
exchange from diagram (a) in Fig.~\ref{fig1}, yielding \cite{Maekawa:2011vs,deVries:2012ab,Bsaisou:2014oka,Gnech:2019dod}
\beqa
W_{-, \; 1\pi}^{{\rm I} \; (Q^{-1})} &=& \frac{g_A \bar g_0}{2 F_\pi}
\frac{1}{q^2 + M_\pi^2}\,, \nn
V_{-, \; 1\pi}^{{\rm II} \; (Q^{-1})} &=&\frac{g_A \bar g_2}{2 F_\pi}
\frac{1}{q^2 + M_\pi^2}\,,\nn
V_{-, \; 1\pi}^{{\rm III} \; (Q^{-1})} &=& V_{+, \; 1\pi}^{{\rm IV} \;
  (Q^{-1})} \; = \; \frac{g_A \bar g_1}{4 F_\pi}
\frac{1}{q^2 + M_\pi^2}\,.
\eeqa
The first corrections to the PVTV potential emerge at order $Q^0$ from one-loop
diagrams (b) and (c) in Fig.~\ref{fig1}. The latter graph yields
vanishing contribution, while the long-range potential generated by diagram (b)
has the form \cite{deVries:2012ab,Bsaisou:2014oka,Gnech:2019dod}
\beq
\label{NLOTPENonRen}
V_{-, \; 1\pi + 2 \pi}^{{\rm III} \, (Q^{0})} = V_{+, \; 1\pi + 2 \pi}^{{\rm IV} \,
  (Q^{0})}  =  - \frac{5 g_A^3 \Delta_3 m_N}{64 \pi F_\pi^3 (q
  ^2 + M_\pi^2)} \big[ ( q^2 + 2 M_\pi^2 ) A(q) + M_\pi \big] \,,
\eeq
with the loop function $A (q)$ given by
\beq
A(q) = \frac{1}{2 q} \arctan \bigg( \frac{q}{2 M_\pi} \bigg)\,.
\eeq
Here and in what follows, the ultraviolet divergent loop contributions to the PVTV
potentials are regularized using dimensional regularization. Clearly, cutoff
regularization would lead to the same result up
to contributions that vanish in the infinite-cutoff limit. The
expression in Eq.~(\ref{NLOTPENonRen}) possesses a pion pole and
therefore corresponds to both one- and two-pion exchange.
As done in the parity conserving sector, we employ the on-shell
renormalization scheme for PVTV pion-nucleon coupling constants
$\bar g_{1,2,3}$ by defining their renormalized values from the residues of
the corresponding form factors. This way, the pion-pole contribution
in Eq.~(\ref{NLOTPENonRen}) gets absorbed into a (finite) renormalization of the
LEC $\bar g_1$,
yielding the two-pion exchange potential
\beq
\label{NLOTPERen}
V_{-, \; 2 \pi}^{{\rm III} \, (Q^{0})} = V_{+, \; 2 \pi}^{{\rm IV} \,
  (Q^{0})}  =  - \frac{5 g_A^3 \Delta_3 m_N}{64 \pi F_\pi^3 (q
  ^2 + M_\pi^2)} \bigg[( q^2 + 2 M_\pi^2 ) A(q) - \frac{M_\pi}{2} \mathrm{arctanh} \bigg( \frac{1}{2} \bigg) \bigg] \,.
\eeq
The only remaining NLO contribution in the delta-less formulation of
chiral EFT emerges from diagram (d) in Fig.~\ref{fig1}. The
corresponding three-nucleon force has the form \cite{Bsaisou:2014oka,Gnech:2019dod}  
\beq
\label{NLO3NF}
V_{\rm 3N}^{(Q^{0})}=-\frac{ig_A^3\Delta_3m_N}{4
  F_\pi^3}  \frac{(\vec \sigma_1\cdot\vec
  q_1)(\vec \sigma_2\cdot\vec q _2)(\vec \sigma_3 \cdot \vec q_3)}{(q_1^2+M_\pi^2)(q_2^2+M_\pi^2)(q_3^2+M_\pi^2)} \big[(\fet \tau_2\cdot\fet \tau_3)\tau_1^3+(\fet
\tau_1\cdot\fet \tau_3)\tau_2^3+(\fet \tau_1\cdot\fet \tau_2)\tau_3^3
\big]
\,,
\eeq
where $\vec q_i = \vec p_i \,' - \vec p_i$ is the momentum transfer of
the nucleon $i$, and $\vec q_1 + \vec q_2 + \vec q_3 = 0$ because of
momentum conservation.  

Next, at N$^2$LO, the contributions to the PVTV 2N
potential from diagrams (e), (f), (g) and (h) in Fig.~\ref{fig1} have
the form  
\beqa
\label{NNLOTPENonRen}
W_{-, \; 2 \pi}^{{\rm I} \, (Q^{1})} &=& - \frac{g_A (2 \bar g_0 +\bar g_2)}{32
  \pi^2 F_\pi^3 (q^2 + 4 M_\pi^2)}
\big[ (8 g_A^2 -4) M_\pi^2 + (3 g_A^2 -1 ) q^2 \big] L(q)\,,\nn
V_{-, \; 2 \pi}^{{\rm II} \, (Q^{1})} &=&  \frac{g_A \bar g_2}{32
  \pi^2 F_\pi^3 (q^2 + 4 M_\pi^2)}
\big[ (8 g_A^2 -4) M_\pi^2 + (3 g_A^2 -1 ) q^2 \big] L(q)\,,\nn
V_{-, \; 1\pi +2 \pi}^{{\rm III} \, (Q^{1})} &=& V_{+, \; 1\pi+2 \pi}^{{\rm IV} \,
  (Q^{1})}  = \frac{5 g_A \Delta_3 m_N}{192 \pi^2 F_\pi^3}
\bigg[\frac{3 M_\pi^2}{q^2 + M_\pi^2} (8 c_1 - c_2 - 2 c_3 ) - (c_2 +
6 c_3 ) \bigg] L (q)\,,
\eeqa
where only terms nonpolynomial in momenta are listed. 
The loop function $L (q)$ is given by
\beq
L(q) = \frac{s}{q} \ln \bigg( \frac{s+q}{2 M_\pi} \bigg)\,,
\eeq
with $s = \sqrt{q^2 + 4 M_\pi^2}$. The $\bar g_0$-contribution to the
potential $W_{-, \; 2
  \pi}^{{\rm I} \, (Q^{1})}$ agrees with the one given in
Ref.~\cite{Gnech:2019dod}. On the other hand, the $\bar g_2$-contribution
to $W_{-, \; 2
  \pi}^{{\rm I} \, (Q^{1})}$
 and the potential  $V_{-, \; 2 \pi}^{{\rm II} \,
   (Q^{1})}$ differ from the corresponding expressions
 in Ref.~\cite{Gnech:2019dod} by the factors of $3/4$ and $1/2$,
 respectively. These discrepancies originate from an incorrect factor in
 front of the second term in Eq.~(B16) of Ref.~\cite{Gnech:2019dod}
 obtained by adding the expressions in Eqs.~(B14) and (B15). As for
$V_{-, \; 1\pi +2 \pi}^{{\rm III} \, (Q^{1})} $ and $V_{-, \; 1\pi +2
  \pi}^{{\rm IV} \, (Q^{1})}$, the term $\propto c_1$ agrees with
the one given in Ref.~\cite{Gnech:2019dod}, while those $\propto c_2$
and $\propto c_3$ differ from the expressions given in that paper. 
Up to equation (B28) of Ref.~\cite{Gnech:2019dod}, we actually found the same
expressions as given in that paper.
The discrepancy seems to originate from evaluating the integral in
front of $c_2+c_3$ in Eq.~(B28) of that work. 
The ultraviolet divergences generated by
loop integrals are absorbed into renormalization of $\bar g_1$ and of five
N$^2$LO PVTV contact interactions, whose explicit form can be found in
Ref.~\cite{Gnech:2019dod}. Similarly to the NLO potentials in Eq.~(\ref{NLOTPENonRen}), the
pion-pole contribution to  $V_{-, \; 1\pi +2 \pi}^{{\rm III} \, (Q^{1})} $ and $V_{-, \; 1\pi +2
  \pi}^{{\rm IV} \, (Q^{1})}$ in Eq.~(\ref{NNLOTPENonRen}) gets absorbed into
renormalization of the LEC $\bar g_1$, leading to the two-pion exchange
potential
\beq
\label{eq:nnlo_nodelta_mom}
V_{-, \; 2 \pi}^{{\rm III} \, (Q^{1})}  = V_{+, \; 2 \pi}^{{\rm IV} \,
  (Q^{1})}  = \frac{5 g_A \Delta_3 m_N}{192 \pi^2 F_\pi^3}
\bigg[\frac{3 M_\pi^2}{q^2 + M_\pi^2} (8 c_1 - c_2 - 2 c_3 ) \bigg(
L(q) - \frac{\pi}{2 \sqrt{3}} \bigg)- (c_2 +
6 c_3 )L (q)  \bigg] \,.
\eeq

We now turn to the delta-isobar contributions.
The NLO contribution is generated by diagram (j) in Fig.~\ref{fig1} and can be expressed as
\beq
\label{eq:nlo_delta_mom}
V_{-, \; 1\pi+2 \pi , \; \Delta}^{{\rm III}\, (Q^{0})} =V_{+, \;
  1\pi+2 \pi , \; \Delta}^{{\rm IV}\, (Q^{0})}= -\frac{5\Delta_3 g_A h_A^2 m_N}{36 \pi^2 F_\pi^3}\frac{\Delta}{M_\pi^2+q^2}\left[(2M_\pi^2+q^2-2\Delta^2)D(q)-L(q)\right]\,,
\eeq
where $\Delta$ denotes the mass splitting $\Delta = m_\Delta - m_N$
and  we have introduced the loop function
\beq
D(q)=\frac{1}{\Delta}\int_{2M_\pi}^\infty d\mu \frac{1}{\mu^2+q^2}\arctan\left(\frac{\sqrt{\mu^2-4M_\pi}}{2\Delta}\right)\,.
\eeq
Subtracting out the pion-pole contribution that renormalizes the LEC $\bar
g_1$, we obtain the corresponding two-pion exchange potential in the
form
\beq
\label{eq:nlo_delta_mom_TPEP}
V_{-, \; 2 \pi , \; \Delta}^{{\rm III}\, (Q^{0})} =V_{+, \; 2 \pi , \;
  \Delta}^{{\rm IV}\, (Q^{0})}= -\frac{5\Delta_3 g_A h_A^2 m_N}{36  \pi^2 F_\pi^3
 }\frac{\Delta}{M_\pi^2+q^2}\left[(2M_\pi^2+q^2-2\Delta^2)D(q)-(M_\pi^2
  - 2 \Delta^2) D(i M_\pi) -
  L(q) + \frac{\pi}{2\sqrt{3}}\right]\,.
\eeq

At N$^2$LO, we find for the 2N-diagrams (k) and (l) in Fig.~\ref{fig1} 
\beqa
\label{eq:nnlo_delta_mom}
V_{-, \; 2 \pi , \; \Delta}^{{\rm I}\, (Q^{1})} &=& -\frac{g_A h_A^2(3\bar g_0+\bar g_2)}{18 \pi F_\pi^3 \Delta}(2 M_\pi^2+q^2) A(q)\,,\nn
W_{-, \; 2 \pi , \; \Delta}^{{\rm I}\, (Q^{1})} &=&- \frac{(2\bar
  g_0+\bar g_2)g_A h_A^2}{36\pi^2F_\pi^3}\left[ (2M_\pi^2+q^2-2\Delta^2)D(q) -L(q)\right]\,,\nn
V_{-, \; 2 \pi , \; \Delta}^{{\rm II}\, (Q^{1})} &=& \frac{\bar g_2g_A
h_A^2}{36\pi^2 F_\pi^3}\left[ (2M_\pi^2+q^2-2\Delta^2)D(q) -L(q) \right]\,,\nn
V_{-, \; 2 \pi , \; \Delta}^{{\rm III}\, (Q^{1})} &=& V_{+, \; 2 \pi ,
  \; \Delta}^{{\rm IV}\, (Q^{1})} = -\frac{g_A h_A^2 \bar g_1}{36 \pi F_\pi^3\Delta} (2M_\pi^2+q^2) A(q)\,.
\eeqa
Finally, the diagram (m) generates the 3N-potential given by
\beqa
\label{Delta3NF}
V_{3N}^{(Q^{1})} &=& \frac{4g_A h_A^2}{9F_\pi^3\Delta} \frac{\vec q_1\cdot\vec q_2}{
  (q_1^2+M_\pi^2)(q_3^2+M_\pi^2)} \bigg\{
\frac{\bar g_1}{2}\Big[\big(\vec\sigma_1\cdot\vec
  q_1-\vec\sigma_3\cdot \vec q_3 \big)
  \big(\tau_1^3+\tau_3^3\big)+\big(\vec\sigma_1\cdot\vec
  q_1+\vec\sigma_3\cdot \vec
  q_3\big)\big(\tau_1^3-\tau_3^3\big)\Big]
  \nn
  &+& \bar g_2 \, \tau_1^3\tau_3^3
  \, \big(\vec\sigma_1\cdot\vec q_1- \vec\sigma_3\cdot\vec q_3 \big) \bigg\} \; +\; \mathrm{5
    \; permutations}\,.
\eeqa

As a consistency check of the calculated $\Delta$ contributions to the
PVTV 2N potential, we have looked at their representation in terms of
resonance saturation of LECs in the delta-less approach. Using the
$1/\Delta$ expansion of the loop function $D(q)$,  
\beq
D(q)=-\frac{1}{2\Delta^2}L(q)-\frac{1}{24\Delta^4}(4M_\pi^2+q^2)L(q)+\mathcal{O}(\Delta^{-6}),
\eeq
one observes that the $\mathcal{O} (\Delta^{-1})$ contributions
stemming from the $1/\Delta$ expansion of $V_{-, \; 2 \pi, \;
  \Delta}^{{\rm III}\,({Q^0})}$ and $V_{-, \; 2 \pi, \;
  \Delta}^{{\rm IV}\,({Q^0})}$ in Eq.~(\ref{eq:nlo_delta_mom_TPEP})
are reproduced by setting the LECs $c_i$ in
Eq.~(\ref{eq:nnlo_nodelta_mom}) to their $\Delta$-resonance-saturation
values \cite{Bernard:1996gq}
\beq
\label{LECsResSat}
c_1=0,\quad c_2=-c_3=\frac{4h_A^2}{9\Delta}\,.
\eeq
To verify the consistency of the N$^2$LO-$\Delta$ contributions given in
Eq.~(\ref{eq:nnlo_delta_mom}), we have worked out the triangle diagram
of type (g) in Fig.~\ref{fig1}, where the Weinberg-Tomozawa vertex is
replaced by the subleading $\pi \pi NN$ vertex proportional to the
LECs $c_i$. This yields the N$^3$LO contributions of the form
\beqa
\label{eq:nnnlo_saturation}
V_{-,\; 2 \pi }^{{\rm I}\,({Q^2})} &=& \frac{g_A(3\bar g_0+\bar
  g_2)}{8\pi F_\pi^3}\big[M_\pi^2(4c_1-2c_3)-q^2c_3 \big]A(q)\,, \nn
V_{-,\; 2 \pi }^{{\rm III}\,({Q^2})} = V_{+,\; 2 \pi }^{{\rm
    IV}\,({Q^2})} &=& \frac{g_A\bar g_1}{16\pi F_\pi^3}\big[M_\pi^2(4c_1-2c_3)-q^2c_3\big]A(q)\,.
\eeqa
Again, it is easy to see that these expressions coincide with the
$\mathcal{O} (\Delta^{-1})$ terms resulting from the
$1/\Delta$-expansion of Eq.~(\ref{eq:nnlo_delta_mom}) when using the
values of $c_i$'s specified in Eq.~(\ref{LECsResSat}). We have also
verified the resonance-saturation picture for the three-nucleon force
given in Eq.~(\ref{Delta3NF}).

\section{Large-distance behavior of the PVTV 2N potentials}
\label{sec:CoordSpace}

The strength of the calculated long-range potentials can be read off from the
corresponding coordinate-space expressions. Following the notation in
Eq.~(\ref{NotationMom}), we define the profile functions 
$\tilde V_-^{\rm I} (r)$, $\tilde W_-^{\rm I} (r)$,   $\tilde V_-^{\rm II} (r)$,   $\tilde V_-^{\rm
  III} (r)$ and $\tilde V_+^{\rm IV} (r)$ via
\beq
\label{NotationCoord}
\tilde V (\vec r \, ) = \big[ \tilde V_-^{\rm I} + \tilde W_-^{\rm I} \fet \tau_1 \cdot \fet \tau_2  +
\tilde V_-^{\rm II}  \tau_1^3
\tau_2^3 + \tilde V_-^{\rm III} (\tau_1^3 + \tau_2^3) \big] \,  (\hat r \cdot
\vec \sigma_1 - \hat r \cdot \vec \sigma_2 ) + \tilde V_+^{\rm IV} (\tau_1^3
- \tau_2^3) \,  
 (\hat r \cdot
\vec \sigma_1 + \hat r \cdot \vec \sigma_2 ) \,,
\eeq
where $\vec r$ is the relative distance between the nucleons, $r
\equiv | \vec r \,|$ and $\hat
r \equiv \vec r / r$.
The PVTV one-pion exchange potential has the form
\beqa
\tilde W_{-, \; 1\pi}^{{\rm I}\,({Q^{-1}})} (r) &=&  - \frac{g_A \bar
  g_0}{8 \pi F_\pi} \frac{e^{-x}}{r^2} (1 + x)\,, \nn
\tilde V_{-, \; 1\pi}^{{\rm II}\,({Q^{-1}})} (r) &=&  - \frac{g_A \bar
  g_2}{8 \pi F_\pi} \frac{e^{-x}}{r^2} (1 + x)\,,\nn
\tilde V_{-, \; 1\pi}^{{\rm III} \; (Q^{-1})} (r) &=& \tilde V_{+, \; 1\pi}^{{\rm IV} \;
  (Q^{-1})} (r) \; = \; - \frac{g_A \bar
  g_1}{16 \pi F_\pi} \frac{e^{-x}}{r^2} (1 + x)\,,
\eeqa
where we have introduced a dimensionless variable $x = M_\pi r$.  

The Fourier transform of the two-pion exchange
contributions can be facilitated using their spectral representation \cite{Kaiser:1997mw},
which yields the large-distance behavior 
\beq
\label{VrSpectr}
\tilde{X}_\pm (r) =-\frac{1}{2\pi^2r^2}\int_{2M_\pi}^{\infty} d\mu\mu
e^{-\mu r}\, (1+\mu r) \, \mathrm{Im} X_\pm (-i\mu)\,,
\eeq
in terms of the spectral functions $\mathrm{Im} X_\pm (-i\mu)$ with
$X$ staying for $V$ or $W$.
Hence, to calculate the coordinate expressions, we need the imaginary parts of the loop functions. By analytic continuation one obtains
\beqa
\label{ImParts}
\mathrm{Im} L(-i\mu) &=& -\frac{\pi}{2} \frac{\sqrt{\mu^2-4M_\pi^2}}{\mu}\,,\nn
\mathrm{Im} A(-i\mu) &=& \frac{\pi}{4\mu}\,,\nn
\mathrm{Im} D(-i\mu) &=& \frac{\pi}{2\Delta\mu} \arctan{\frac{\sqrt{\mu^2-4M_\pi^2}}{2\Delta}}\,.
\eeqa
Notice that the coordinate-space expressions given below become
meaningless at short distances of $r$ well below  $M_\pi^{-1}$, which
is outside of the convergence range of the chiral expansion.
When calculating nuclear observables by solving the Schr\"odinger equation, this unphysical
short-range behavior is removed by a regulator. 

For the leading two-pion exchange contributions given in Eq.~(\ref{NLOTPERen}),
we obtain
\beqa
\tilde V_{-, \; 2 \pi}^{{\rm III} \, (Q^{0})} (r)
&=&\tilde V_{+, \; 2 \pi}^{{\rm IV} \, (Q^{0})} (r) \; = \; \frac{5
  g_A^3 \Delta_3 m_N}{1024 \pi^2 F_\pi^3} \frac{1}{r^3} \Big[ x (x-1)
e^x {\rm Ei} (-3x) + 4(1+x) e^{-2x} + x(1+x) e^{-x}  {\rm Ei} (-x) \Big]\,,
\eeqa
where ${\rm Ei} (z)$ is the exponential integral function given by
${\rm Ei} (z) = - \int_{z}^\infty e^{-t}/t dt$. 
For the subleading two-pion exchange contributions specified in
Eq.~(\ref{NNLOTPENonRen}), we have only succeeded to obtain analytical results for
$\tilde W_{-, \; 2 \pi}^{{\rm I} \, (Q^{1})} (r)$ and $\tilde V_{-, \;
  2 \pi}^{{\rm II} \, (Q^{0})} (r)$:
\beqa
\tilde{W}_{-,\;2\pi}^{{\rm I}\,(Q^{1})}&=&-\frac{(2\bar g_0+\bar g_2)g_AM_\pi}{64\pi^3F_\pi^3 r^3}\Big\{2x(4g_A^2-1)K_0(2x)+\big[g_A^2(9+4x^2)-3\big]K_1(2x)\Big\} \,, \nn
\tilde{V}_{-,\;2\pi}^{{\rm II}\,(Q^{1})}&=&\frac{\bar
  g_2g_AM_\pi}{64\pi^3F_\pi^3r^3}\Big\{2x(4g_A^2-1)
K_0(2x)+\big[g_A^2(9+4x^2)-3\big] K_1(2x)\Big\}\,, 
\eeqa
where $K_n (z)$ refer to the modified Bessel functions of the second
kind. The remaining potentials $\tilde{V}_{-,\;2\pi}^{{\rm
    III}\,(Q^{1})} (r)$ and $\tilde{V}_{+,\;2\pi}^{{\rm
    IV}\,(Q^{1})} (r)$ can be obtained numerically by performing the
spectral integral in Eq.~(\ref{VrSpectr}) using
Eqs.~(\ref{NNLOTPENonRen}) and  (\ref{ImParts}). 

For the two-pion exchange potentials mediated by the intermediate
delta excitations, compact analytical expressions are only
available for $\tilde{V}_{-,\;2\pi, \; \Delta}^{{\rm
    I}\,(Q^{1})} (r)$, $\tilde{V}_{-,\;2\pi, \; \Delta}^{{\rm
    III}\,(Q^{1})} (r)$ and $\tilde{V}_{+,\;2\pi, \; \Delta}^{{\rm
    IV}\,(Q^{1})} (r)$,
\beqa
\label{eq:nnlo_delta_coord}
\tilde V_{-, \; 2 \pi , \; \Delta}^{{\rm I}\, (Q^{1})} (r)&=&-\frac{(3\bar
  g_0+\bar g_2)g_A h_A^2}{36 \pi^2 F_\pi^3\Delta}\frac{e^{-2x}}{r^5}(1+x)\big(2+x(2+x)\big),\\
\tilde V_{-, \; 2 \pi , \; \Delta}^{{\rm III}\, (Q^{1})} (r)&=& \tilde
V_{+, \; 2 \pi , \; \Delta}^{{\rm IV}\, (Q^{1})} (r) \; =\; -\frac{\bar g_1g_A h_A^2}{72\pi^2 F_\pi^3 \Delta}\frac{e^{-2x}}{r^5}(1+x)\big(2+x(2+x)\big)\,,
\eeqa
while the remaining potentials have to be calculated numerically using
Eqs.~(\ref{eq:nlo_delta_mom_TPEP}), (\ref{eq:nnlo_delta_mom}), (\ref{VrSpectr}) and (\ref{ImParts}).  

We are now in the position to present numerical results  and discuss the impact of treating the delta
isobar as an explicit degree of freedom on the PVTV 2N potentials. Here and in what follows, we use the following values for the various constants:
 $M_\pi=138$~MeV, $F_\pi = 92.4$~MeV, $m_N = 939$~MeV, $\Delta =
 293$~MeV. Further, we use the effective value for the axial coupling
 constant of the nucleon, $g_A = 1.29$, which takes into account the
 Goldberger-Treiman discrepancy. For the pion-nucleon-delta  coupling
 constant $h_A$, we adopt the value of $h_A = 1.34$ fixed from the
 width of the delta resonance. For the LECs $c_i$, we employ the
 values $c_1=-0.74$~GeV$^{-1}$, $c_2=1.81$~GeV$^{-1}$ and
 $c_3=-3.61$~GeV$^{-1}$ taken from the order-$Q^2$ heavy-baryon-NN fit
 of Ref.~\cite{Siemens:2016jwj}. When performing the calculations in
 the formulation with explicit delta isobars, we subtract from these
 values the delta  contributions specified in Eq.~(\ref{LECsResSat}) and use
 $c_1=-0.74$~GeV$^{-1}$, $c_2=-0.91$~GeV$^{-1}$ and
 $c_3=-0.89$~GeV$^{-1}$. This leaves us with the only unknown
 (dimensionless) LECs $\bar g_0$, $\bar g_1$, $\bar g_2$ and
 $\Delta_3$ associated with the PVTV vertices. 
The dependence of the obtained
contributions to the one- and two-pion exchange 2N potentials on these
PVTV LECs is summarized in Table \ref{Tab1}. 
\begin{table*}
   \begin{ruledtabular}
    \begin{tabular*}{\textwidth}{@{\extracolsep{\fill}}rrr|cccc|ccc}
&&& &&&&&\\[-12pt]
&&& LO ($V_{1 \pi}$)  & NLO ($V_{2 \pi}$) & N$^2$LO ($V_{2 \pi}$) &&
                                                                 NLO-$\Delta$
                                                                     ($V_{2
                                                                     \pi}$)
      & N$^2$LO-$\Delta$ ($V_{2 \pi}$)&\\
 &&& &&&&&\\[-12pt]     
      \hline
      &$V_-^{\rm I}$ &&--- & --- & ---& & --- & $3 \bar g_0 + \bar g_2$&\\
      &$W_-^{\rm I}$ && $\bar g_0$ & --- & $2 \bar g_0 + \bar g_2$ & & --- & $2 \bar g_0 + \bar
                                                g_2$&\\
        &$V_-^{\rm II}$ &&$\bar g_2$ & --- & $\bar g_2$ & & --- & $\bar g_2$&\\
        &$V_-^{\rm III}$ && $\bar g_1$ & $\Delta_3$ & $\Delta_3$& & $\Delta_3$ & $\bar g_1$&\\   
  \end{tabular*}
  \caption{Low-energy constants accompanying various PVTV vertices
    that govern the long-range contributions to the PVTV 2N force up
    to N$^2$LO in the chiral EFT expansion. NLO and N$^2$LO refer to
    the two-pion exchange contributions in the delta-less formulation,
    which are generated by diagrams (b), (c) and (e)-(h) of
    Fig.~\ref{fig1}. NLO-$\Delta$ and N$^2$LO-$\Delta$ refer to the
    additional contributions in the delta-full framework stemming from
    diagrams (j), (k) and (l).}
  \label{Tab1}
  \end{ruledtabular} 
\end{table*}
Notice that while the NLO-$\Delta$ contribution to the two-pion
exchange is implicitly taken into account in the delta-less framework
through resonance saturation of the LECs $c_i$ entering the N$^2$LO
potential, the N$^2$LO-$\Delta$ terms go beyond the N$^2$LO accuracy level of the
delta-less formulation of chiral EFT. It is thus interesting to look at the 
strength of the resulting potentials, which may give hints about their
phenomenological importance when analyzing PVTV signals in nuclear systems.  

\begin{figure}[tb]
\includegraphics[width=0.45\textwidth]{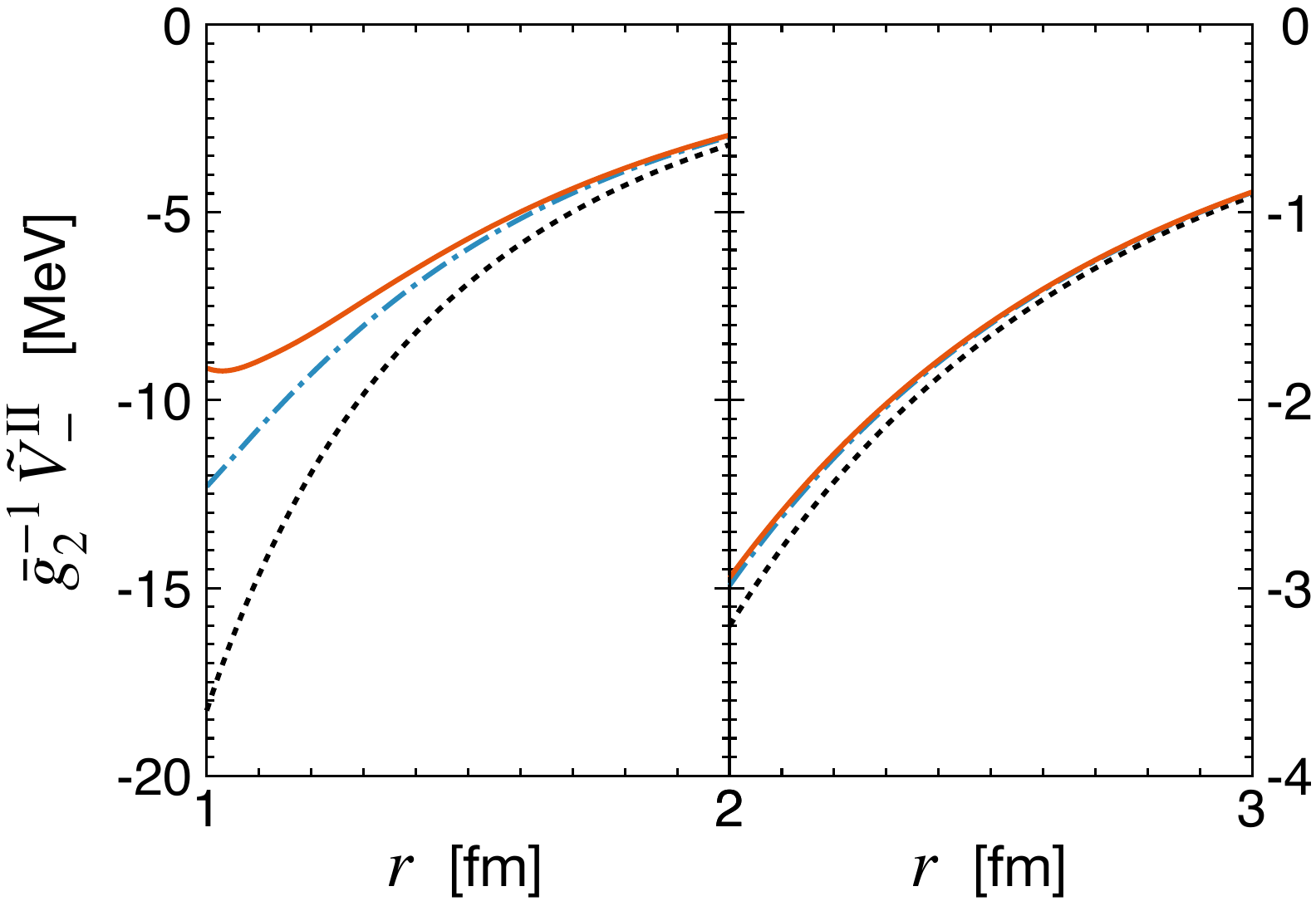}\\
  \caption{Chiral EFT expansion of the PVTV 2N potential
    $\tilde{V}_{-}^{{\rm II}} (r)$ as a function of the distance $r$
    between the nucleons. Black dotted lines show the LO result due to
  the one-pion exchange, while blue dashed-dotted and red solid lines
  refer to the complete N$^2$LO results in the delta-less and delta-full
formulations of chiral EFT, respectively. Notice that there are no NLO
contributions to $\tilde{V}_{-}^{{\rm II}} (r)$.}  
\label{fig2}
\end{figure}

In Fig.~\ref{fig2}, we compare the chiral expansion of the potential
$\tilde V_-^{\rm II} (r)$ in the delta-less and delta-full formulations
of chiral EFT. Notice that this structure  does not receive any
corrections at order $Q^0$ (i.e., at NLO). Clearly, the long-range part of the interaction
is completely dominated by the one-pion exchange, but the effects of
the two-pion exchange become visible at distances below about $
2$~fm. For the case at hand, the delta-resonance contribution at
N$^2$LO turns out to be smaller in magnitude  than the nucleonic one at the same
order and it, in fact, becomes negligibly small beyond $r \sim 1.5$~fm. 

\begin{figure}[tb]
\includegraphics[width=\textwidth]{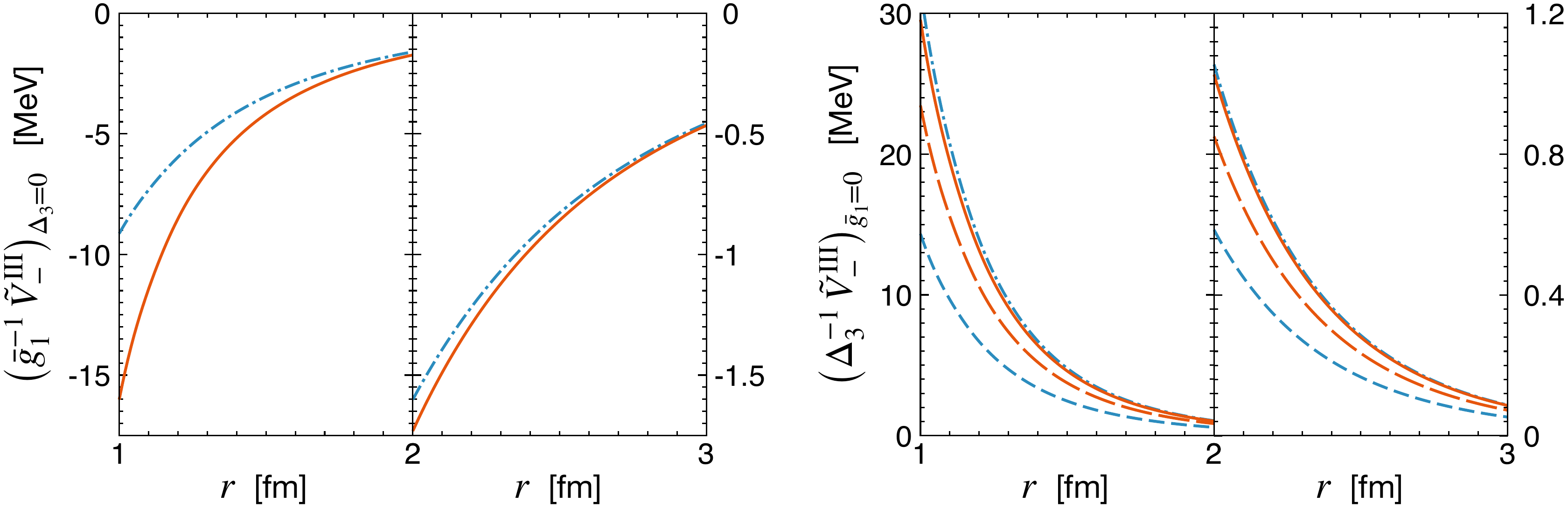}\\
  \caption{Chiral EFT expansion of the PVTV 2N potential
    $\tilde{V}_{-}^{{\rm III}} (r)$ as a function of the distance $r$
    between the nucleons. Left and right panels show the results for
    the contributions proportional to the PVTV LECs $\bar g_1$ and
    $\Delta_3$, respectively. Blue short-dashed and red long-dashed
    lines show the complete NLO results in the delta-less and delta-full
formulations of chiral EFT, respectively. For the remaining notation
see Fig.~\ref{fig2}. }
\label{fig3}
\end{figure}

Next, in Fig.~\ref{fig3}, we compare the results obtained for the PVTV
potential $\tilde V_-^{\rm III} (r)$  using the delta-less and
delta-full formulations of chiral EFT. Since the contributions at
different orders depend on two unknown PVTV LECs $\bar g_1$ and $\Delta_3$, we
have decided to show the corresponding potentials separately. The
$\bar g_1$-contribution is dominated by the one-pion exchange, which
does not receive any corrections at NLO and N$^2$LO in the delta-less
formulation. When the delta-isobar is treated as an explicit degree of
freedom, the $\bar g_1$-part of the potential $\tilde V_-^{\rm III}
(r)$ receives a correction at N$^2$LO, which appears to be quite
sizable at distances $r \lesssim 1.5$~fm. The rather large
contribution of the delta resonance is consistent with the enhancement
by a factor of $\pi$, as one can see from
Eq.~(\ref{eq:nnlo_delta_mom}). The potential $\tilde V_-^{\rm III}
(r)$ receives the two-pion exchange contributions at NLO and
N$^2$LO\footnote{As  explained in section \ref{sec:MomSpace}, diagrams (b)
  and (j) of Fig.~\ref{fig1} also generate contributions to the
  one-pion exchange potential which, however, only lead to
  renormalization of the LEC $\bar g_1$ and are thus not shown in the
  right panel of Fig.~\ref{fig3}.}. In the delta-less framework,
the N$^2$LO correction to $\tilde V_-^{\rm III} (r)$ is large, being
numerically comparable to the nominally dominant NLO contribution. The convergence of the
chiral EFT expansion is considerably improved in the delta-full
framework, where the dominant NLO-$\Delta$ result $\tilde V_{-, \; 2 \pi , \;
  \Delta}^{{\rm III}\, (Q^{0})} (r)$ already takes into
account a significant part of the N$^2$LO correction. In particular,
it includes the part of the N$^2$LO potential $\tilde V_{-, \; 2
  \pi}^{{\rm III} \, (Q^{1})} (r)$, which is  driven by the
delta-resonance-saturation values of the LECs $c_i$ quoted in
Eq.~(\ref{LECsResSat}). The potential $\tilde V_-^{\rm III} (r)$ does
not receive any N$^2$LO contribution $\propto \Delta_3$ from the
intermediate delta excitation, so that the N$^2$LO correction is given
by the same expression in both the delta-less and delta-full
formulations. However, in the approach with explicit delta-isobar, the
LECs $c_{2,3}$ are considerably smaller in magnitude, so that the
N$^2$LO-$\Delta$ correction has a more natural size. The final results
at N$^2$LO, however, appear to be rather similar in both
frameworks. This may be viewed as an indication that effects of the
delta-isobar beyond the leading contributions governed by the resonance
saturation of $c_i$'s are small.  

\begin{figure}[tb]
\includegraphics[width=\textwidth]{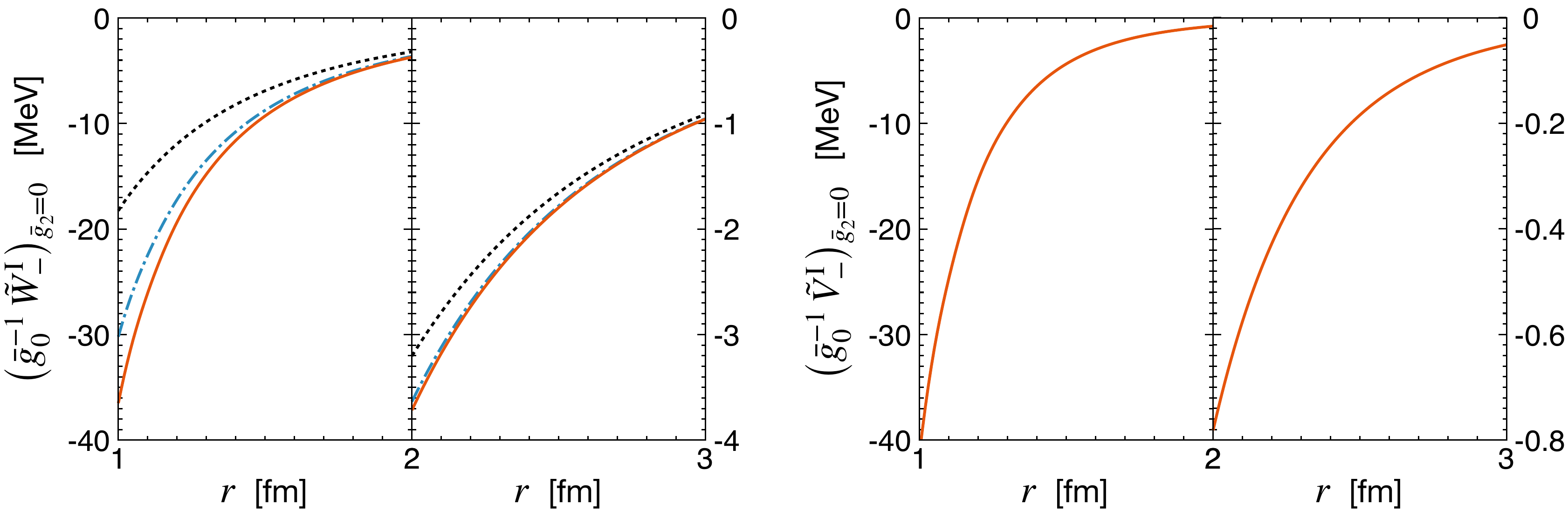}\\
  \caption{Chiral EFT expansion of the PVTV 2N potentials
    $\tilde{W}_{-}^{{\rm I}} (r)$ (left panel) and
    $\tilde{V}_{-}^{{\rm I}} (r)$ (right panel)  as a function of the distance $r$
    between the nucleons. In both cases, the PVTV LEC $\bar g_2$ is
    set to zero. For notation
see Figs.~\ref{fig2} and \ref{fig3}. }
\label{fig4}
\end{figure}

Finally, we turn to the isospin-invariant PVTV potentials $\tilde
V_-^{\rm I} (r)$ and $\tilde W_-^{\rm I} (r)$. As shown in Table
\ref{Tab1}, these potentials are driven by different linear
combinations of the unknown PVTV coupling constants $\bar g_0$ and $\bar g_2$,
which complicates the comparison of their strengths at different EFT
orders. However, in most scenarios of CP violation in and
beyond the Standard Model, the coupling $\bar g_2$ is strongly
suppressed relative to $\bar g_0$ \cite{deVries:2012ab}, see Table I of
Ref.~\cite{deVries:2020iea}. In particular, in the scenario with
strong CP violation 
generated by the QCD $\theta$-term, one even has $\Delta_3, \bar g_1, \bar
g_2 \ll \bar g_0 \sim \mathcal{O} (1)$.  In the following, we
therefore set   $\bar g_2 = 0$ for the sake of comparison of the
delta-full and delta-less results. Our predictions for the potentials 
 $\tilde W_-^{\rm I} (r)$ and $\tilde
V_-^{\rm I} (r)$ are shown in Fig.~\ref{fig4}. The isovector potential
$\tilde W_-^{\rm I} (r)$ receives the contribution from the one-pion
exchange at LO, which dominates its large-distance behavior. 
The N$^2$LO two-pion exchange potential $\tilde W_{-, \; 2 \pi}^{{\rm I}
  \, (Q^{1})} (r)$ provides a sizable correction at $r \lesssim 2$~fm,
while the corresponding delta-resonance contribution $\tilde W_{-, \; 2 \pi ,
  \; \Delta}^{{\rm I}\, (Q^{1})} (r)$ is fairly small, especially at
large distances.  The most interesting finding of our work is the
PVTV isoscalar potential $\tilde V_-^{\rm I} (r)$, which is
shown in the right panel of Fig.~\ref{fig4}. It receives no
contribution from the one-pion exchange and two-pion exchange up-to-and-including
N$^2$LO in the delta-less formulation. On the other hand, we find a
very strong two-pion exchange contribution at N$^2$LO-$\Delta$ generated by diagrams
(k) and (j) in Fig.~\ref{fig1} with an intermediate
delta-excitation. In the delta-less framework, these contributions
appear first at N$^3$LO from triangle diagrams with one
insertion of the subleading pion-nucleon vertices proportional to
$c_i$'s, see Eq.~(\ref{eq:nnnlo_saturation}). The strong enhancement of $\tilde V_{-, \; 2 \pi , \; \Delta}^{{\rm I}\,
  (Q^{1})} (r)$ relativ to the expectations based on the power
counting is similar to what is observed for the isoscalar central PCTC
two-pion exchange potential \cite{Epelbaum:2024gfg}. As one can see from Eq.~(\ref{eq:nnlo_delta_mom})
or Eq.~(\ref{eq:nnnlo_saturation}), the enhancement originates from
a combination of factors including the smallness of the delta-nucleon
mass splitting $\Delta$, the enhancement by a factor of $\pi$ and 
a large numerical prefactor. As a result, the potential $\tilde V_{-, \; 2 \pi , \; \Delta}^{{\rm I}\,
  (Q^{1})} (r)$ appears to be even stronger than all of the NLO PVTV two-pion exchange
components we have calculated. Thus, one may expect from this novel
contribution to significantly affect the predictions for PVTV nuclear observables.

\section{Summary and conclusions}
\label{sec:Summary}

In this paper, we have derived the contributions to the long-range 
parity-violating time-reversal-violating nuclear forces up through
N$^2$LO using the formulation of chiral EFT with explicit delta-isobar
degrees of freedom.  The main results of our study can be summarized
as follows:
\begin{itemize}
\item We have discussed the PVTV effective Lagrangian for pions,
  nucleons and the delta-isobar and argued that PVTV $\pi N
  \Delta$-vertices must involve at least one derivative, thus being
  suppressed relative to the leading derivative-less pion and
  pion-nucleon vertices. Accordingly, no additional unknown
  coupling constants appear in the delta-contributions to the  PVTV nuclear
  forces up to N$^2$LO.  Using the delta-full formulation of chiral EFT
  thus allows one to resum the dominant contributions beyond N$^2$LO of
  the delta-less framework in a parameter-free way, leading to an
  improved convergence of the EFT expansion. 
\item
 Our renormalized expressions for the long-range two- and
 three-nucleon PVTV forces in the delta-less chiral EFT framework mostly agree with the ones given in the
 literature, but we also found some differences for the N$^2$LO
 two-pion exchange potentials as compared with Ref.~\cite{Gnech:2019dod}. 
\item
  We have worked out the delta contributions to the long-range two- and
  three-nucleon potentials at NLO-$\Delta$ and N$^2$LO-$\Delta$ using
  the small-scale expansion scheme by counting $\Delta \sim M_\pi$.
  We have verified that all obtained contributions are consistent with
  the delta-resonance saturation picture, i.e., the dominant terms
  after performing the $1/\Delta$-expansion coinside with the
  one-order-higher delta-less expressions upon using the values of the
  LECs $c_i$ quoted in Eq.~(\ref{LECsResSat}).
\item
  By analyzing the potentials in coordinate space, we found
  indications of an improved convergence of the small-scale expansion
  compared to the delta-less formulation of chiral EFT for the PVTV
  potential $\tilde V_-^{\rm III} (r)$. As the most interesting result
  of this study, we found a strong isospin-invariant PVTV two-pion
  exchange potential $\tilde V_-^{\rm I} (r)$ driven by the intermediate
  delta excitation, which is completely missing in the delta-less
  framework at the N$^2$LO accuracy level. 
\end{itemize}  
The analytical expressions for various potentials we obtained are well
suited for applications to few- and many-nucleon systems upon
introducing an appropriate regulator as done e.g.~in Ref.~\cite{Reinert:2017usi}. It
would be interesting to see phenomenological implications of our
findings on the relationships between the PVTV couplings $\Delta_3$,
$\bar g_i$ and PVTV few-nucleon observables such as, e.g., the
electric dipole moments of light nuclei, see Ref.~\cite{deVries:2020iea} for a
review.

\section*{Acknowledgments}

This work was supported in part by the MKW NRW
under the funding code NW21-024-A, by
DFG and NSFC through funds provided to the Sino-German CRC 110
``Symmetries and the Emergence of Structure in QCD'' (NSFC Grant
No. 11621131001, DFG Project-ID 196253076 - TRR 110),
by ERC  NuclearTheory (grant No. 885150) and by the EU Horizon 2020 research and
innovation programme (STRONG-2020, grant agreement No. 824093).


\bibliographystyle{apsrev4-1}
\bibliography{refs}

\begin{thebibliography}{34}%
\makeatletter
\providecommand \@ifxundefined [1]{%
 \@ifx{#1\undefined}
}%
\providecommand \@ifnum [1]{%
 \ifnum #1\expandafter \@firstoftwo
 \else \expandafter \@secondoftwo
 \fi
}%
\providecommand \@ifx [1]{%
 \ifx #1\expandafter \@firstoftwo
 \else \expandafter \@secondoftwo
 \fi
}%
\providecommand \natexlab [1]{#1}%
\providecommand \enquote  [1]{``#1''}%
\providecommand \bibnamefont  [1]{#1}%
\providecommand \bibfnamefont [1]{#1}%
\providecommand \citenamefont [1]{#1}%
\providecommand \href@noop [0]{\@secondoftwo}%
\providecommand \href [0]{\begingroup \@sanitize@url \@href}%
\providecommand \@href[1]{\@@startlink{#1}\@@href}%
\providecommand \@@href[1]{\endgroup#1\@@endlink}%
\providecommand \@sanitize@url [0]{\catcode `\\12\catcode `\$12\catcode
  `\&12\catcode `\#12\catcode `\^12\catcode `\_12\catcode `\%12\relax}%
\providecommand \@@startlink[1]{}%
\providecommand \@@endlink[0]{}%
\providecommand \url  [0]{\begingroup\@sanitize@url \@url }%
\providecommand \@url [1]{\endgroup\@href {#1}{\urlprefix }}%
\providecommand \urlprefix  [0]{URL }%
\providecommand \Eprint [0]{\href }%
\providecommand \doibase [0]{http://dx.doi.org/}%
\providecommand \selectlanguage [0]{\@gobble}%
\providecommand \bibinfo  [0]{\@secondoftwo}%
\providecommand \bibfield  [0]{\@secondoftwo}%
\providecommand \translation [1]{[#1]}%
\providecommand \BibitemOpen [0]{}%
\providecommand \bibitemStop [0]{}%
\providecommand \bibitemNoStop [0]{.\EOS\space}%
\providecommand \EOS [0]{\spacefactor3000\relax}%
\providecommand \BibitemShut  [1]{\csname bibitem#1\endcsname}%
\let\auto@bib@innerbib\@empty
\bibitem [{\citenamefont {Mereghetti}\ \emph {et~al.}(2010)\citenamefont
  {Mereghetti}, \citenamefont {Hockings},\ and\ \citenamefont {van
  Kolck}}]{Mereghetti:2010tp}%
  \BibitemOpen
  \bibfield  {author} {\bibinfo {author} {\bibfnamefont {E.}~\bibnamefont
  {Mereghetti}}, \bibinfo {author} {\bibfnamefont {W.~H.}\ \bibnamefont
  {Hockings}}, \ and\ \bibinfo {author} {\bibfnamefont {U.}~\bibnamefont {van
  Kolck}},\ }\href {\doibase 10.1016/j.aop.2010.03.005} {\bibfield  {journal}
  {\bibinfo  {journal} {Annals Phys.}\ }\textbf {\bibinfo {volume} {325}},\
  \bibinfo {pages} {2363} (\bibinfo {year} {2010})},\ \Eprint
  {http://arxiv.org/abs/1002.2391} {arXiv:1002.2391 [hep-ph]} \BibitemShut
  {NoStop}%
\bibitem [{\citenamefont {de~Vries}\ \emph {et~al.}(2013)\citenamefont
  {de~Vries}, \citenamefont {Mereghetti}, \citenamefont {Timmermans},\ and\
  \citenamefont {van Kolck}}]{deVries:2012ab}%
  \BibitemOpen
  \bibfield  {author} {\bibinfo {author} {\bibfnamefont {J.}~\bibnamefont
  {de~Vries}}, \bibinfo {author} {\bibfnamefont {E.}~\bibnamefont
  {Mereghetti}}, \bibinfo {author} {\bibfnamefont {R.~G.~E.}\ \bibnamefont
  {Timmermans}}, \ and\ \bibinfo {author} {\bibfnamefont {U.}~\bibnamefont {van
  Kolck}},\ }\href {\doibase 10.1016/j.aop.2013.05.022} {\bibfield  {journal}
  {\bibinfo  {journal} {Annals Phys.}\ }\textbf {\bibinfo {volume} {338}},\
  \bibinfo {pages} {50} (\bibinfo {year} {2013})},\ \Eprint
  {http://arxiv.org/abs/1212.0990} {arXiv:1212.0990 [hep-ph]} \BibitemShut
  {NoStop}%
\bibitem [{\citenamefont {Bsaisou}\ \emph
  {et~al.}(2015{\natexlab{a}})\citenamefont {Bsaisou}, \citenamefont
  {de~Vries}, \citenamefont {Hanhart}, \citenamefont {Liebig}, \citenamefont
  {Mei{\ss}ner}, \citenamefont {Minossi}, \citenamefont {Nogga},\ and\
  \citenamefont {Wirzba}}]{Bsaisou:2014zwa}%
  \BibitemOpen
  \bibfield  {author} {\bibinfo {author} {\bibfnamefont {J.}~\bibnamefont
  {Bsaisou}}, \bibinfo {author} {\bibfnamefont {J.}~\bibnamefont {de~Vries}},
  \bibinfo {author} {\bibfnamefont {C.}~\bibnamefont {Hanhart}}, \bibinfo
  {author} {\bibfnamefont {S.}~\bibnamefont {Liebig}}, \bibinfo {author}
  {\bibfnamefont {U.-G.}\ \bibnamefont {Mei{\ss}ner}}, \bibinfo {author}
  {\bibfnamefont {D.}~\bibnamefont {Minossi}}, \bibinfo {author} {\bibfnamefont
  {A.}~\bibnamefont {Nogga}}, \ and\ \bibinfo {author} {\bibfnamefont
  {A.}~\bibnamefont {Wirzba}},\ }\href {\doibase 10.1007/JHEP03(2015)104}
  {\bibfield  {journal} {\bibinfo  {journal} {JHEP}\ }\textbf {\bibinfo
  {volume} {03}},\ \bibinfo {pages} {104} (\bibinfo {year}
  {2015}{\natexlab{a}})},\ \bibinfo {note} {[Erratum: JHEP 05, 083 (2015)]},\
  \Eprint {http://arxiv.org/abs/1411.5804} {arXiv:1411.5804 [hep-ph]}
  \BibitemShut {NoStop}%
\bibitem [{\citenamefont {Bsaisou}\ \emph
  {et~al.}(2015{\natexlab{b}})\citenamefont {Bsaisou}, \citenamefont
  {Mei\ss{}ner}, \citenamefont {Nogga},\ and\ \citenamefont
  {Wirzba}}]{Bsaisou:2014oka}%
  \BibitemOpen
  \bibfield  {author} {\bibinfo {author} {\bibfnamefont {J.}~\bibnamefont
  {Bsaisou}}, \bibinfo {author} {\bibfnamefont {U.-G.}\ \bibnamefont
  {Mei\ss{}ner}}, \bibinfo {author} {\bibfnamefont {A.}~\bibnamefont {Nogga}},
  \ and\ \bibinfo {author} {\bibfnamefont {A.}~\bibnamefont {Wirzba}},\ }\href
  {\doibase 10.1016/j.aop.2015.04.031} {\bibfield  {journal} {\bibinfo
  {journal} {Annals Phys.}\ }\textbf {\bibinfo {volume} {359}},\ \bibinfo
  {pages} {317} (\bibinfo {year} {2015}{\natexlab{b}})},\ \Eprint
  {http://arxiv.org/abs/1412.5471} {arXiv:1412.5471 [hep-ph]} \BibitemShut
  {NoStop}%
\bibitem [{\citenamefont {de~Vries}\ \emph {et~al.}(2020)\citenamefont
  {de~Vries}, \citenamefont {Epelbaum}, \citenamefont {Girlanda}, \citenamefont
  {Gnech}, \citenamefont {Mereghetti},\ and\ \citenamefont
  {Viviani}}]{deVries:2020iea}%
  \BibitemOpen
  \bibfield  {author} {\bibinfo {author} {\bibfnamefont {J.}~\bibnamefont
  {de~Vries}}, \bibinfo {author} {\bibfnamefont {E.}~\bibnamefont {Epelbaum}},
  \bibinfo {author} {\bibfnamefont {L.}~\bibnamefont {Girlanda}}, \bibinfo
  {author} {\bibfnamefont {A.}~\bibnamefont {Gnech}}, \bibinfo {author}
  {\bibfnamefont {E.}~\bibnamefont {Mereghetti}}, \ and\ \bibinfo {author}
  {\bibfnamefont {M.}~\bibnamefont {Viviani}},\ }\href {\doibase
  10.3389/fphy.2020.00218} {\bibfield  {journal} {\bibinfo  {journal} {Front.
  in Phys.}\ }\textbf {\bibinfo {volume} {8}},\ \bibinfo {pages} {218}
  (\bibinfo {year} {2020})},\ \Eprint {http://arxiv.org/abs/2001.09050}
  {arXiv:2001.09050 [nucl-th]} \BibitemShut {NoStop}%
\bibitem [{\citenamefont {Maekawa}\ \emph {et~al.}(2011)\citenamefont
  {Maekawa}, \citenamefont {Mereghetti}, \citenamefont {de~Vries},\ and\
  \citenamefont {van Kolck}}]{Maekawa:2011vs}%
  \BibitemOpen
  \bibfield  {author} {\bibinfo {author} {\bibfnamefont {C.~M.}\ \bibnamefont
  {Maekawa}}, \bibinfo {author} {\bibfnamefont {E.}~\bibnamefont {Mereghetti}},
  \bibinfo {author} {\bibfnamefont {J.}~\bibnamefont {de~Vries}}, \ and\
  \bibinfo {author} {\bibfnamefont {U.}~\bibnamefont {van Kolck}},\ }\href
  {\doibase 10.1016/j.nuclphysa.2011.09.020} {\bibfield  {journal} {\bibinfo
  {journal} {Nucl. Phys. A}\ }\textbf {\bibinfo {volume} {872}},\ \bibinfo
  {pages} {117} (\bibinfo {year} {2011})},\ \Eprint
  {http://arxiv.org/abs/1106.6119} {arXiv:1106.6119 [nucl-th]} \BibitemShut
  {NoStop}%
\bibitem [{\citenamefont {Bsaisou}\ \emph {et~al.}(2013)\citenamefont
  {Bsaisou}, \citenamefont {Hanhart}, \citenamefont {Liebig}, \citenamefont
  {Mei{\ss}ner}, \citenamefont {Nogga},\ and\ \citenamefont
  {Wirzba}}]{Bsaisou:2012rg}%
  \BibitemOpen
  \bibfield  {author} {\bibinfo {author} {\bibfnamefont {J.}~\bibnamefont
  {Bsaisou}}, \bibinfo {author} {\bibfnamefont {C.}~\bibnamefont {Hanhart}},
  \bibinfo {author} {\bibfnamefont {S.}~\bibnamefont {Liebig}}, \bibinfo
  {author} {\bibfnamefont {U.-G.}\ \bibnamefont {Mei{\ss}ner}}, \bibinfo
  {author} {\bibfnamefont {A.}~\bibnamefont {Nogga}}, \ and\ \bibinfo {author}
  {\bibfnamefont {A.}~\bibnamefont {Wirzba}},\ }\href {\doibase
  10.1140/epja/i2013-13031-x} {\bibfield  {journal} {\bibinfo  {journal} {Eur.
  Phys. J. A}\ }\textbf {\bibinfo {volume} {49}},\ \bibinfo {pages} {31}
  (\bibinfo {year} {2013})},\ \Eprint {http://arxiv.org/abs/1209.6306}
  {arXiv:1209.6306 [hep-ph]} \BibitemShut {NoStop}%
\bibitem [{\citenamefont {Gnech}\ and\ \citenamefont
  {Viviani}(2020)}]{Gnech:2019dod}%
  \BibitemOpen
  \bibfield  {author} {\bibinfo {author} {\bibfnamefont {A.}~\bibnamefont
  {Gnech}}\ and\ \bibinfo {author} {\bibfnamefont {M.}~\bibnamefont
  {Viviani}},\ }\href {\doibase 10.1103/PhysRevC.101.024004} {\bibfield
  {journal} {\bibinfo  {journal} {Phys. Rev. C}\ }\textbf {\bibinfo {volume}
  {101}},\ \bibinfo {pages} {024004} (\bibinfo {year} {2020})},\ \Eprint
  {http://arxiv.org/abs/1906.09021} {arXiv:1906.09021 [nucl-th]} \BibitemShut
  {NoStop}%
\bibitem [{\citenamefont {de~Vries}\ \emph {et~al.}(2011)\citenamefont
  {de~Vries}, \citenamefont {Higa}, \citenamefont {Liu}, \citenamefont
  {Mereghetti}, \citenamefont {Stetcu}, \citenamefont {Timmermans},\ and\
  \citenamefont {van Kolck}}]{deVries:2011an}%
  \BibitemOpen
  \bibfield  {author} {\bibinfo {author} {\bibfnamefont {J.}~\bibnamefont
  {de~Vries}}, \bibinfo {author} {\bibfnamefont {R.}~\bibnamefont {Higa}},
  \bibinfo {author} {\bibfnamefont {C.~P.}\ \bibnamefont {Liu}}, \bibinfo
  {author} {\bibfnamefont {E.}~\bibnamefont {Mereghetti}}, \bibinfo {author}
  {\bibfnamefont {I.}~\bibnamefont {Stetcu}}, \bibinfo {author} {\bibfnamefont
  {R.~G.~E.}\ \bibnamefont {Timmermans}}, \ and\ \bibinfo {author}
  {\bibfnamefont {U.}~\bibnamefont {van Kolck}},\ }\href {\doibase
  10.1103/PhysRevC.84.065501} {\bibfield  {journal} {\bibinfo  {journal} {Phys.
  Rev. C}\ }\textbf {\bibinfo {volume} {84}},\ \bibinfo {pages} {065501}
  (\bibinfo {year} {2011})},\ \Eprint {http://arxiv.org/abs/1109.3604}
  {arXiv:1109.3604 [hep-ph]} \BibitemShut {NoStop}%
\bibitem [{\citenamefont {Ordonez}\ \emph {et~al.}(1996)\citenamefont
  {Ordonez}, \citenamefont {Ray},\ and\ \citenamefont {van
  Kolck}}]{Ordonez:1995rz}%
  \BibitemOpen
  \bibfield  {author} {\bibinfo {author} {\bibfnamefont {C.}~\bibnamefont
  {Ordonez}}, \bibinfo {author} {\bibfnamefont {L.}~\bibnamefont {Ray}}, \ and\
  \bibinfo {author} {\bibfnamefont {U.}~\bibnamefont {van Kolck}},\ }\href
  {\doibase 10.1103/PhysRevC.53.2086} {\bibfield  {journal} {\bibinfo
  {journal} {Phys. Rev. C}\ }\textbf {\bibinfo {volume} {53}},\ \bibinfo
  {pages} {2086} (\bibinfo {year} {1996})},\ \Eprint
  {http://arxiv.org/abs/hep-ph/9511380} {arXiv:hep-ph/9511380} \BibitemShut
  {NoStop}%
\bibitem [{\citenamefont {Kaiser}\ \emph {et~al.}(1998)\citenamefont {Kaiser},
  \citenamefont {Gerstendorfer},\ and\ \citenamefont {Weise}}]{Kaiser:1998wa}%
  \BibitemOpen
  \bibfield  {author} {\bibinfo {author} {\bibfnamefont {N.}~\bibnamefont
  {Kaiser}}, \bibinfo {author} {\bibfnamefont {S.}~\bibnamefont
  {Gerstendorfer}}, \ and\ \bibinfo {author} {\bibfnamefont {W.}~\bibnamefont
  {Weise}},\ }\href {\doibase 10.1016/S0375-9474(98)00234-6} {\bibfield
  {journal} {\bibinfo  {journal} {Nucl. Phys. A}\ }\textbf {\bibinfo {volume}
  {637}},\ \bibinfo {pages} {395} (\bibinfo {year} {1998})},\ \Eprint
  {http://arxiv.org/abs/nucl-th/9802071} {arXiv:nucl-th/9802071} \BibitemShut
  {NoStop}%
\bibitem [{\citenamefont {Krebs}\ \emph {et~al.}(2007)\citenamefont {Krebs},
  \citenamefont {Epelbaum},\ and\ \citenamefont {Mei{\ss}ner}}]{Krebs:2007rh}%
  \BibitemOpen
  \bibfield  {author} {\bibinfo {author} {\bibfnamefont {H.}~\bibnamefont
  {Krebs}}, \bibinfo {author} {\bibfnamefont {E.}~\bibnamefont {Epelbaum}}, \
  and\ \bibinfo {author} {\bibfnamefont {U.-G.}\ \bibnamefont {Mei{\ss}ner}},\
  }\href {\doibase 10.1140/epja/i2007-10372-y} {\bibfield  {journal} {\bibinfo
  {journal} {Eur. Phys. J. A}\ }\textbf {\bibinfo {volume} {32}},\ \bibinfo
  {pages} {127} (\bibinfo {year} {2007})},\ \Eprint
  {http://arxiv.org/abs/nucl-th/0703087} {arXiv:nucl-th/0703087} \BibitemShut
  {NoStop}%
\bibitem [{\citenamefont {Epelbaum}\ \emph
  {et~al.}(2008{\natexlab{a}})\citenamefont {Epelbaum}, \citenamefont {Krebs},\
  and\ \citenamefont {Mei{\ss}ner}}]{Epelbaum:2007sq}%
  \BibitemOpen
  \bibfield  {author} {\bibinfo {author} {\bibfnamefont {E.}~\bibnamefont
  {Epelbaum}}, \bibinfo {author} {\bibfnamefont {H.}~\bibnamefont {Krebs}}, \
  and\ \bibinfo {author} {\bibfnamefont {U.-G.}\ \bibnamefont {Mei{\ss}ner}},\
  }\href {\doibase 10.1016/j.nuclphysa.2008.02.305} {\bibfield  {journal}
  {\bibinfo  {journal} {Nucl. Phys. A}\ }\textbf {\bibinfo {volume} {806}},\
  \bibinfo {pages} {65} (\bibinfo {year} {2008}{\natexlab{a}})},\ \Eprint
  {http://arxiv.org/abs/0712.1969} {arXiv:0712.1969 [nucl-th]} \BibitemShut
  {NoStop}%
\bibitem [{\citenamefont {Epelbaum}\ \emph
  {et~al.}(2008{\natexlab{b}})\citenamefont {Epelbaum}, \citenamefont {Krebs},\
  and\ \citenamefont {Mei{\ss}ner}}]{Epelbaum:2008td}%
  \BibitemOpen
  \bibfield  {author} {\bibinfo {author} {\bibfnamefont {E.}~\bibnamefont
  {Epelbaum}}, \bibinfo {author} {\bibfnamefont {H.}~\bibnamefont {Krebs}}, \
  and\ \bibinfo {author} {\bibfnamefont {U.-G.}\ \bibnamefont {Mei{\ss}ner}},\
  }\href {\doibase 10.1103/PhysRevC.77.034006} {\bibfield  {journal} {\bibinfo
  {journal} {Phys. Rev. C}\ }\textbf {\bibinfo {volume} {77}},\ \bibinfo
  {pages} {034006} (\bibinfo {year} {2008}{\natexlab{b}})},\ \Eprint
  {http://arxiv.org/abs/0801.1299} {arXiv:0801.1299 [nucl-th]} \BibitemShut
  {NoStop}%
\bibitem [{\citenamefont {Krebs}\ \emph {et~al.}(2018)\citenamefont {Krebs},
  \citenamefont {Gasparyan},\ and\ \citenamefont {Epelbaum}}]{Krebs:2018jkc}%
  \BibitemOpen
  \bibfield  {author} {\bibinfo {author} {\bibfnamefont {H.}~\bibnamefont
  {Krebs}}, \bibinfo {author} {\bibfnamefont {A.~M.}\ \bibnamefont
  {Gasparyan}}, \ and\ \bibinfo {author} {\bibfnamefont {E.}~\bibnamefont
  {Epelbaum}},\ }\href {\doibase 10.1103/PhysRevC.98.014003} {\bibfield
  {journal} {\bibinfo  {journal} {Phys. Rev. C}\ }\textbf {\bibinfo {volume}
  {98}},\ \bibinfo {pages} {014003} (\bibinfo {year} {2018})},\ \Eprint
  {http://arxiv.org/abs/1803.09613} {arXiv:1803.09613 [nucl-th]} \BibitemShut
  {NoStop}%
\bibitem [{\citenamefont {Gasser}\ and\ \citenamefont
  {Leutwyler}(1984)}]{Gasser:1983yg}%
  \BibitemOpen
  \bibfield  {author} {\bibinfo {author} {\bibfnamefont {J.}~\bibnamefont
  {Gasser}}\ and\ \bibinfo {author} {\bibfnamefont {H.}~\bibnamefont
  {Leutwyler}},\ }\href {\doibase 10.1016/0003-4916(84)90242-2} {\bibfield
  {journal} {\bibinfo  {journal} {Annals Phys.}\ }\textbf {\bibinfo {volume}
  {158}},\ \bibinfo {pages} {142} (\bibinfo {year} {1984})}\BibitemShut
  {NoStop}%
\bibitem [{\citenamefont {Bernard}\ \emph {et~al.}(1995)\citenamefont
  {Bernard}, \citenamefont {Kaiser},\ and\ \citenamefont
  {Mei{\ss}ner}}]{Bernard:1995dp}%
  \BibitemOpen
  \bibfield  {author} {\bibinfo {author} {\bibfnamefont {V.}~\bibnamefont
  {Bernard}}, \bibinfo {author} {\bibfnamefont {N.}~\bibnamefont {Kaiser}}, \
  and\ \bibinfo {author} {\bibfnamefont {U.-G.}\ \bibnamefont {Mei{\ss}ner}},\
  }\href {\doibase 10.1142/S0218301395000092} {\bibfield  {journal} {\bibinfo
  {journal} {Int. J. Mod. Phys. E}\ }\textbf {\bibinfo {volume} {4}},\ \bibinfo
  {pages} {193} (\bibinfo {year} {1995})},\ \Eprint
  {http://arxiv.org/abs/hep-ph/9501384} {arXiv:hep-ph/9501384} \BibitemShut
  {NoStop}%
\bibitem [{\citenamefont {Fettes}\ \emph {et~al.}(2000)\citenamefont {Fettes},
  \citenamefont {Mei{\ss}ner}, \citenamefont {Mojzis},\ and\ \citenamefont
  {Steininger}}]{Fettes:2000gb}%
  \BibitemOpen
  \bibfield  {author} {\bibinfo {author} {\bibfnamefont {N.}~\bibnamefont
  {Fettes}}, \bibinfo {author} {\bibfnamefont {U.-G.}\ \bibnamefont
  {Mei{\ss}ner}}, \bibinfo {author} {\bibfnamefont {M.}~\bibnamefont {Mojzis}},
  \ and\ \bibinfo {author} {\bibfnamefont {S.}~\bibnamefont {Steininger}},\
  }\href {\doibase 10.1006/aphy.2000.6059} {\bibfield  {journal} {\bibinfo
  {journal} {Annals Phys.}\ }\textbf {\bibinfo {volume} {283}},\ \bibinfo
  {pages} {273} (\bibinfo {year} {2000})},\ \bibinfo {note} {[Erratum: Annals
  Phys. 288, 249--250 (2001)]},\ \Eprint {http://arxiv.org/abs/hep-ph/0001308}
  {arXiv:hep-ph/0001308} \BibitemShut {NoStop}%
\bibitem [{\citenamefont {Hemmert}\ \emph {et~al.}(1998)\citenamefont
  {Hemmert}, \citenamefont {Holstein},\ and\ \citenamefont
  {Kambor}}]{Hemmert:1997ye}%
  \BibitemOpen
  \bibfield  {author} {\bibinfo {author} {\bibfnamefont {T.~R.}\ \bibnamefont
  {Hemmert}}, \bibinfo {author} {\bibfnamefont {B.~R.}\ \bibnamefont
  {Holstein}}, \ and\ \bibinfo {author} {\bibfnamefont {J.}~\bibnamefont
  {Kambor}},\ }\href {\doibase 10.1088/0954-3899/24/10/003} {\bibfield
  {journal} {\bibinfo  {journal} {J. Phys. G}\ }\textbf {\bibinfo {volume}
  {24}},\ \bibinfo {pages} {1831} (\bibinfo {year} {1998})},\ \Eprint
  {http://arxiv.org/abs/hep-ph/9712496} {arXiv:hep-ph/9712496} \BibitemShut
  {NoStop}%
\bibitem [{\citenamefont {Weinberg}(1990)}]{Weinberg:1990rz}%
  \BibitemOpen
  \bibfield  {author} {\bibinfo {author} {\bibfnamefont {S.}~\bibnamefont
  {Weinberg}},\ }\href {\doibase 10.1016/0370-2693(90)90938-3} {\bibfield
  {journal} {\bibinfo  {journal} {Phys. Lett. B}\ }\textbf {\bibinfo {volume}
  {251}},\ \bibinfo {pages} {288} (\bibinfo {year} {1990})}\BibitemShut
  {NoStop}%
\bibitem [{\citenamefont {Epelbaum}(2007)}]{Epelbaum:2007us}%
  \BibitemOpen
  \bibfield  {author} {\bibinfo {author} {\bibfnamefont {E.}~\bibnamefont
  {Epelbaum}},\ }\href {\doibase 10.1140/epja/i2007-10496-0} {\bibfield
  {journal} {\bibinfo  {journal} {Eur. Phys. J. A}\ }\textbf {\bibinfo {volume}
  {34}},\ \bibinfo {pages} {197} (\bibinfo {year} {2007})},\ \Eprint
  {http://arxiv.org/abs/0710.4250} {arXiv:0710.4250 [nucl-th]} \BibitemShut
  {NoStop}%
\bibitem [{\citenamefont {Epelbaum}\ \emph {et~al.}(2009)\citenamefont
  {Epelbaum}, \citenamefont {Hammer},\ and\ \citenamefont
  {Mei{\ss}ner}}]{Epelbaum:2008ga}%
  \BibitemOpen
  \bibfield  {author} {\bibinfo {author} {\bibfnamefont {E.}~\bibnamefont
  {Epelbaum}}, \bibinfo {author} {\bibfnamefont {H.-W.}\ \bibnamefont
  {Hammer}}, \ and\ \bibinfo {author} {\bibfnamefont {U.-G.}\ \bibnamefont
  {Mei{\ss}ner}},\ }\href {\doibase 10.1103/RevModPhys.81.1773} {\bibfield
  {journal} {\bibinfo  {journal} {Rev. Mod. Phys.}\ }\textbf {\bibinfo {volume}
  {81}},\ \bibinfo {pages} {1773} (\bibinfo {year} {2009})},\ \Eprint
  {http://arxiv.org/abs/0811.1338} {arXiv:0811.1338 [nucl-th]} \BibitemShut
  {NoStop}%
\bibitem [{\citenamefont {Kaiser}(2007)}]{Kaiser:2007zzb}%
  \BibitemOpen
  \bibfield  {author} {\bibinfo {author} {\bibfnamefont {N.}~\bibnamefont
  {Kaiser}},\ }\href {\doibase 10.1103/PhysRevC.76.047001} {\bibfield
  {journal} {\bibinfo  {journal} {Phys. Rev. C}\ }\textbf {\bibinfo {volume}
  {76}},\ \bibinfo {pages} {047001} (\bibinfo {year} {2007})},\ \Eprint
  {http://arxiv.org/abs/0711.2233} {arXiv:0711.2233 [nucl-th]} \BibitemShut
  {NoStop}%
\bibitem [{\citenamefont {Kaiser}\ \emph {et~al.}(1997)\citenamefont {Kaiser},
  \citenamefont {Brockmann},\ and\ \citenamefont {Weise}}]{Kaiser:1997mw}%
  \BibitemOpen
  \bibfield  {author} {\bibinfo {author} {\bibfnamefont {N.}~\bibnamefont
  {Kaiser}}, \bibinfo {author} {\bibfnamefont {R.}~\bibnamefont {Brockmann}}, \
  and\ \bibinfo {author} {\bibfnamefont {W.}~\bibnamefont {Weise}},\ }\href
  {\doibase 10.1016/S0375-9474(97)00586-1} {\bibfield  {journal} {\bibinfo
  {journal} {Nucl. Phys. A}\ }\textbf {\bibinfo {volume} {625}},\ \bibinfo
  {pages} {758} (\bibinfo {year} {1997})},\ \Eprint
  {http://arxiv.org/abs/nucl-th/9706045} {arXiv:nucl-th/9706045} \BibitemShut
  {NoStop}%
\bibitem [{\citenamefont {Epelbaum}\ \emph {et~al.}(1998)\citenamefont
  {Epelbaum}, \citenamefont {Gl{\"o}ckle},\ and\ \citenamefont
  {Mei{\ss}ner}}]{Epelbaum:1998ka}%
  \BibitemOpen
  \bibfield  {author} {\bibinfo {author} {\bibfnamefont {E.}~\bibnamefont
  {Epelbaum}}, \bibinfo {author} {\bibfnamefont {W.}~\bibnamefont
  {Gl{\"o}ckle}}, \ and\ \bibinfo {author} {\bibfnamefont {U.-G.}\ \bibnamefont
  {Mei{\ss}ner}},\ }\href {\doibase 10.1016/S0375-9474(98)00220-6} {\bibfield
  {journal} {\bibinfo  {journal} {Nucl. Phys. A}\ }\textbf {\bibinfo {volume}
  {637}},\ \bibinfo {pages} {107} (\bibinfo {year} {1998})},\ \Eprint
  {http://arxiv.org/abs/nucl-th/9801064} {arXiv:nucl-th/9801064} \BibitemShut
  {NoStop}%
\bibitem [{\citenamefont {Krebs}\ and\ \citenamefont
  {Epelbaum}(2023)}]{Krebs:2023ljo}%
  \BibitemOpen
  \bibfield  {author} {\bibinfo {author} {\bibfnamefont {H.}~\bibnamefont
  {Krebs}}\ and\ \bibinfo {author} {\bibfnamefont {E.}~\bibnamefont
  {Epelbaum}},\ }\href@noop {} {\  (\bibinfo {year} {2023})},\ \Eprint
  {http://arxiv.org/abs/2311.10893} {arXiv:2311.10893 [nucl-th]} \BibitemShut
  {NoStop}%
\bibitem [{\citenamefont {Epelbaum}(2006)}]{Epelbaum:2005pn}%
  \BibitemOpen
  \bibfield  {author} {\bibinfo {author} {\bibfnamefont {E.}~\bibnamefont
  {Epelbaum}},\ }\href {\doibase 10.1016/j.ppnp.2005.09.002} {\bibfield
  {journal} {\bibinfo  {journal} {Prog. Part. Nucl. Phys.}\ }\textbf {\bibinfo
  {volume} {57}},\ \bibinfo {pages} {654} (\bibinfo {year} {2006})},\ \Eprint
  {http://arxiv.org/abs/nucl-th/0509032} {arXiv:nucl-th/0509032} \BibitemShut
  {NoStop}%
\bibitem [{\citenamefont {Epelbaum}\ \emph {et~al.}(2020)\citenamefont
  {Epelbaum}, \citenamefont {Krebs},\ and\ \citenamefont
  {Reinert}}]{Epelbaum:2019kcf}%
  \BibitemOpen
  \bibfield  {author} {\bibinfo {author} {\bibfnamefont {E.}~\bibnamefont
  {Epelbaum}}, \bibinfo {author} {\bibfnamefont {H.}~\bibnamefont {Krebs}}, \
  and\ \bibinfo {author} {\bibfnamefont {P.}~\bibnamefont {Reinert}},\ }\href
  {\doibase 10.3389/fphy.2020.00098} {\bibfield  {journal} {\bibinfo  {journal}
  {Front. in Phys.}\ }\textbf {\bibinfo {volume} {8}},\ \bibinfo {pages} {98}
  (\bibinfo {year} {2020})},\ \Eprint {http://arxiv.org/abs/1911.11875}
  {arXiv:1911.11875 [nucl-th]} \BibitemShut {NoStop}%
\bibitem [{\citenamefont {Henley}\ and\ \citenamefont
  {Miller}(1979)}]{HenleyMiller}%
  \BibitemOpen
  \bibfield  {author} {\bibinfo {author} {\bibfnamefont {E.~M.}\ \bibnamefont
  {Henley}}\ and\ \bibinfo {author} {\bibfnamefont {G.~A.}\ \bibnamefont
  {Miller}},\ }in\ \href@noop {} {\emph {\bibinfo {booktitle} {Mesons in
  Nuclei}}},\ \bibinfo {editor} {edited by\ \bibinfo {editor} {\bibfnamefont
  {M.}~\bibnamefont {Rho}}\ and\ \bibinfo {editor} {\bibfnamefont
  {D.}~\bibnamefont {Wilkonson}}}\ (\bibinfo  {publisher} {North-Holland},\
  \bibinfo {address} {Amsterdam},\ \bibinfo {year} {1979})\ p.\ \bibinfo
  {pages} {405}\BibitemShut {NoStop}%
\bibitem [{\citenamefont {Epelbaum}\ \emph {et~al.}(2005)\citenamefont
  {Epelbaum}, \citenamefont {Mei{\ss}ner},\ and\ \citenamefont
  {Palomar}}]{Epelbaum:2004xf}%
  \BibitemOpen
  \bibfield  {author} {\bibinfo {author} {\bibfnamefont {E.}~\bibnamefont
  {Epelbaum}}, \bibinfo {author} {\bibfnamefont {U.-G.}\ \bibnamefont
  {Mei{\ss}ner}}, \ and\ \bibinfo {author} {\bibfnamefont {J.~E.}\ \bibnamefont
  {Palomar}},\ }\href {\doibase 10.1103/PhysRevC.71.024001} {\bibfield
  {journal} {\bibinfo  {journal} {Phys. Rev. C}\ }\textbf {\bibinfo {volume}
  {71}},\ \bibinfo {pages} {024001} (\bibinfo {year} {2005})},\ \Eprint
  {http://arxiv.org/abs/nucl-th/0407037} {arXiv:nucl-th/0407037} \BibitemShut
  {NoStop}%
\bibitem [{\citenamefont {Bernard}\ \emph {et~al.}(1997)\citenamefont
  {Bernard}, \citenamefont {Kaiser},\ and\ \citenamefont
  {Mei{\ss}ner}}]{Bernard:1996gq}%
  \BibitemOpen
  \bibfield  {author} {\bibinfo {author} {\bibfnamefont {V.}~\bibnamefont
  {Bernard}}, \bibinfo {author} {\bibfnamefont {N.}~\bibnamefont {Kaiser}}, \
  and\ \bibinfo {author} {\bibfnamefont {U.-G.}\ \bibnamefont {Mei{\ss}ner}},\
  }\href {\doibase 10.1016/S0375-9474(97)00021-3} {\bibfield  {journal}
  {\bibinfo  {journal} {Nucl. Phys. A}\ }\textbf {\bibinfo {volume} {615}},\
  \bibinfo {pages} {483} (\bibinfo {year} {1997})},\ \Eprint
  {http://arxiv.org/abs/hep-ph/9611253} {arXiv:hep-ph/9611253} \BibitemShut
  {NoStop}%
\bibitem [{\citenamefont {Siemens}\ \emph {et~al.}(2017)\citenamefont
  {Siemens}, \citenamefont {Ruiz~de Elvira}, \citenamefont {Epelbaum},
  \citenamefont {Hoferichter}, \citenamefont {Krebs}, \citenamefont {Kubis},\
  and\ \citenamefont {Mei\ss{}ner}}]{Siemens:2016jwj}%
  \BibitemOpen
  \bibfield  {author} {\bibinfo {author} {\bibfnamefont {D.}~\bibnamefont
  {Siemens}}, \bibinfo {author} {\bibfnamefont {J.}~\bibnamefont {Ruiz~de
  Elvira}}, \bibinfo {author} {\bibfnamefont {E.}~\bibnamefont {Epelbaum}},
  \bibinfo {author} {\bibfnamefont {M.}~\bibnamefont {Hoferichter}}, \bibinfo
  {author} {\bibfnamefont {H.}~\bibnamefont {Krebs}}, \bibinfo {author}
  {\bibfnamefont {B.}~\bibnamefont {Kubis}}, \ and\ \bibinfo {author}
  {\bibfnamefont {U.~G.}\ \bibnamefont {Mei\ss{}ner}},\ }\href {\doibase
  10.1016/j.physletb.2017.04.039} {\bibfield  {journal} {\bibinfo  {journal}
  {Phys. Lett. B}\ }\textbf {\bibinfo {volume} {770}},\ \bibinfo {pages} {27}
  (\bibinfo {year} {2017})},\ \Eprint {http://arxiv.org/abs/1610.08978}
  {arXiv:1610.08978 [nucl-th]} \BibitemShut {NoStop}%
\bibitem [{\citenamefont {Epelbaum}(2024)}]{Epelbaum:2024gfg}%
  \BibitemOpen
  \bibfield  {author} {\bibinfo {author} {\bibfnamefont {E.}~\bibnamefont
  {Epelbaum}},\ }\href {\doibase 10.1007/s00601-024-01918-0} {\bibfield
  {journal} {\bibinfo  {journal} {Few Body Syst.}\ }\textbf {\bibinfo {volume}
  {65}},\ \bibinfo {pages} {39} (\bibinfo {year} {2024})}\BibitemShut {NoStop}%
\bibitem [{\citenamefont {Reinert}\ \emph {et~al.}(2018)\citenamefont
  {Reinert}, \citenamefont {Krebs},\ and\ \citenamefont
  {Epelbaum}}]{Reinert:2017usi}%
  \BibitemOpen
  \bibfield  {author} {\bibinfo {author} {\bibfnamefont {P.}~\bibnamefont
  {Reinert}}, \bibinfo {author} {\bibfnamefont {H.}~\bibnamefont {Krebs}}, \
  and\ \bibinfo {author} {\bibfnamefont {E.}~\bibnamefont {Epelbaum}},\ }\href
  {\doibase 10.1140/epja/i2018-12516-4} {\bibfield  {journal} {\bibinfo
  {journal} {Eur. Phys. J. A}\ }\textbf {\bibinfo {volume} {54}},\ \bibinfo
  {pages} {86} (\bibinfo {year} {2018})},\ \Eprint
  {http://arxiv.org/abs/1711.08821} {arXiv:1711.08821 [nucl-th]} \BibitemShut
  {NoStop}%
\end{thebibliography}%

\end{document}